\documentclass[twocolumn,journal]{IEEEtran}
\usepackage[T1]{fontenc}
\usepackage[latin9]{inputenc}
\usepackage{color}
\usepackage{float}
\usepackage{amsmath}
\usepackage{amssymb}
\usepackage{stmaryrd}
\usepackage{graphicx}
\usepackage{setspace}
\usepackage[unicode=true,
 bookmarks=true,bookmarksnumbered=true,bookmarksopen=true,bookmarksopenlevel=1,
 breaklinks=false,pdfborder={0 0 0},pdfborderstyle={},backref=false,colorlinks=false]
 {hyperref}
\hypersetup{pdftitle={Your Title},
 pdfauthor={Your Name},
 pdfpagelayout=OneColumn, pdfnewwindow=true, pdfstartview=XYZ, plainpages=false}

\makeatletter

\floatstyle{ruled}
\newfloat{algorithm}{tbp}{loa}
\providecommand{\algorithmname}{Algorithm}
\floatname{algorithm}{\protect\algorithmname}

\usepackage[caption=false,font=footnotesize]{subfig}
\usepackage{algorithmic}
\usepackage{cite}

\makeatother

\begin{document}
\title{Joint Visibility Region Detection and Channel Estimation for XL-MIMO
Systems via \\Alternating MAP}
\author{Wenkang~Xu,~An Liu, \IEEEmembership{Senior Member,~IEEE,} Min-jian
Zhao{\normalsize{}, and Giuseppe Caire, }\IEEEmembership{Fellow,~IEEE}{\normalsize{}
}\thanks{Wenkang Xu, An Liu, and Min-jian Zhao are with the College of Information
Science and Electronic Engineering, Zhejiang University, Hangzhou
310027, China (email: anliu@zju.edu.cn).

G. Caire is with the Department of Telecommunication Systems, Technical
University of Berlin, 10587 Berlin, Germany (e-mail: caire@tu-berlin.de).}}
\maketitle
\begin{abstract}
We investigate a joint visibility region (VR) detection and channel
estimation problem in extremely large-scale multiple-input-multiple-output
(XL-MIMO) systems, where near-field propagation and spatial non-stationary
effects exist. In this case, each scatterer can only see a subset
of antennas, i.e., it has a certain VR over the antennas. Because
of the spatial correlation among adjacent sub-arrays, VR of scatterers
exhibits a two-dimensional (2D) clustered sparsity. We design a 2D
Markov prior model to capture such a structured sparsity. Based on
this, a novel alternating maximum a posteriori (MAP) framework is
developed for high-accuracy VR detection and channel estimation. The
alternating MAP framework consists of three basic modules: a channel
estimation module, a VR detection module, and a grid update module.
Specifically, the first module is a low-complexity inverse-free variational
Bayesian inference (IF-VBI) algorithm that avoids the matrix inverse
via minimizing a relaxed Kullback-Leibler (KL) divergence. The second
module is a structured expectation propagation (EP) algorithm which
has the ability to deal with complicated prior information. And the
third module refines polar-domain grid parameters via gradient ascent.
Simulations demonstrate the superiority of the proposed algorithm
in both VR detection and channel estimation.
\end{abstract}


\section{Introduction}

High-frequency extremely large-scale multiple-input-multiple-output
(XL-MIMO) has been widely considered as a key technology for future
6G communications \cite{Rappaport_survey}. With the deployment of
hundreds or even thousands of antennas, XL-MIMO systems can mitigate
many of the challenges faced in traditional massive MIMO systems \cite{Hu_survey},
offering unprecedented data rates and enhancing the overall performance
of wireless networks. Meanwhile, for extremely high-frequency communications,
such as millimeter-wave (mmWave) and terahertz (THz) communications,
the size of high-frequency antennas is small due to the small wavelength
\cite{Elayan_Terahertz}. Therefore, it is natural to integrate XL-MIMO
and high-frequency communication into 6G wireless systems \cite{Rappaport_survey}.

\begin{figure}[t]
\begin{centering}
\includegraphics[width=1\columnwidth]{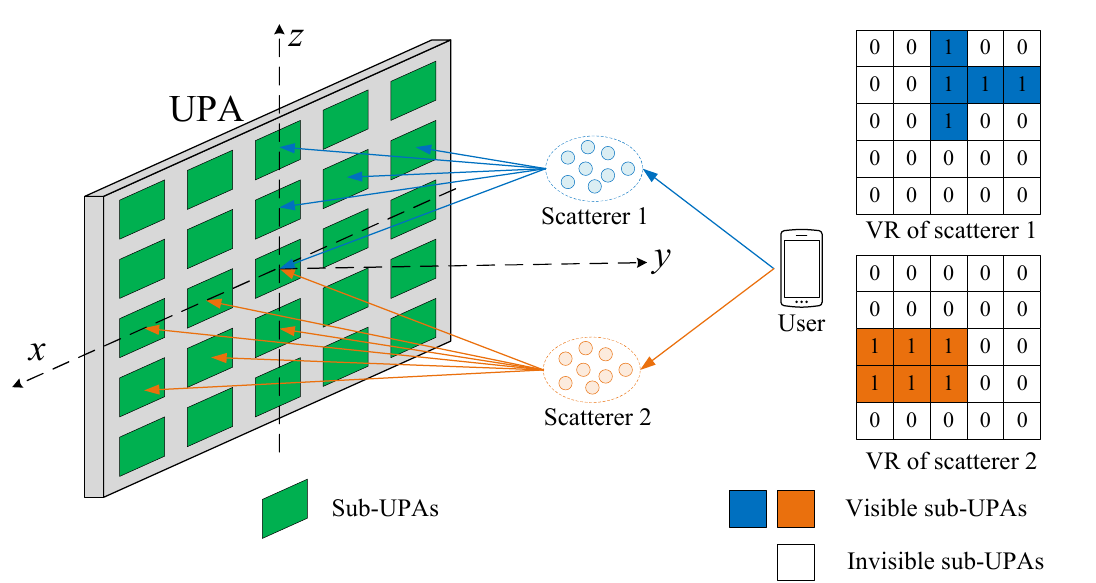}
\par\end{centering}
\caption{\label{fig:Illustration_SNS_VR}Illustration of spatial non-stationary
and VRs in XL-MIMO systems. The VR of scatterers exhibits a clustered
sparsity.}

\end{figure}

However, the use of XL-MIMO and high-frequency bands will lead to
some new issues compared to traditional massive MIMO systems. Above
all, near-field propagation is likely to exist. Given the large antenna
aperture of XL-MIMO and high-frequency, the array will exhibit a large
near-field region \cite{Liu_NF_Tutorial}. When users or scatterers
are inside the Rayleigh region, the array will experience a spherical
wave, which accounts for both angle and distance parameters. Besides,
spatial non-stationary is another issue in high-frequency XL-MIMO
systems. Spatial non-stationary means that different regions of an
extremely large-scale array can see different scatterers \cite{Carvalho_NSN},
making it quite difficult for high-accuracy channel estimation with
low pilot overhead. Fortunately, the measurement in \cite{Yuan_NSN}
verified that sub-channels corresponding to each small-scale sub-array
can be treated as spatial stationary, i.e., each antenna in the same
sub-array can see the same scatterers. In this case, we can focus
on the visibility region (VR) detection for each sub-array instead
of each antenna. Furthermore, there is spatial correlation among adjacent
sub-arrays. As illustrated in Fig. \ref{fig:Illustration_SNS_VR},
adjacent sub-arrays may be visible to the same scatterer with a high
probability \cite{Vincent_sparse_modeling,Tang_sparse_modeling}.
In other words, the VR of a scatterer will concentrate on a few clusters
and the associated VR vector/matrix will exhibit a clustered sparsity.
The above characteristics of XL-MIMO channels should be fully exploited
to enhance the channel estimation performance, especially in the low
signal-to-noise ratio (SNR) regions. Some early attempts at XL-MIMO
channel estimation and visibility region detection are summarized
below.

\textbf{Near-field XL-MIMO channel estimation: }In the case of far-field,
the massive MIMO channel exhibits sparsity in the angular domain.
Many algorithms based on compressive sensing (CS) have been studied
to achieve sparse channel estimation \cite{Lee_FF_CS,LiuAn_CE_Burst_LASSO,LiuAn_CE_Turbo_CS}.
However, in the case of near-field, the steering vector is related
to both the angle and distance parameters of last-hop scatterers.
As a result, the sparse representation for near-field channels is
quite different. In \cite{Cui_polar_grid}, the authors proposed a
polar-domain sparse representation that combined the angle and distance
information simultaneously for near-field channels. Based on the polar-domain
codebook, an off-grid orthogonal matching pursuit (OMP) algorithm
was designed for XL-MIMO channel estimation. In \cite{Zhang_refined_polar},
the authors studied a joint dictionary learning and channel estimation
algorithm to reduce the complexity of the polar-domain dictionary
and enhance the channel estimation performance. The authors in \cite{Yang_3D_polar}
further extended the polar-domain representation from narrow-band
ULA systems to broadband uniform planar array (UPA) systems. In \cite{Lu_CE_UL},
the authors considered a sub-array hybrid precoding architecture and
designed a damped Newtonized OMP algorithm for user localization and
channel estimation. 

\textbf{Joint VR detection and channel estimation: }The above research
only considers the near-field effect, while the spatial non-stationary
issue is not addressed. In \cite{HanYu_VR,HanYu_VR2}, the VR of scatterers
and sub-arrays was studied, and the VR detection was achieved by sub-array
grouping. Inspired by this, several sub-array-wise methods were studied
for addressing the spatial non-stationary issue \cite{HanYu_VR2,Iimori_VR,Chen_NS_CE}.
Specifically, the authors in \cite{HanYu_VR2} designed a joint VR
identification, user localization, and channel estimation scheme with
the aid of reconfigurable intelligent surface (RIS). An expectation-maximization
(EM)-based bilinear Bayesian inference algorithm were proposed in
\cite{Iimori_VR} for joint VR detection and channel estimation. In
\cite{Chen_NS_CE}, the authors exploited the time-domain relevance
of non-stationary effect to improve the performance. Generally speaking,
these sub-array-wise methods first estimate each sub-channel independently,
and then combine the estimated sub-channels into the whole channel.
Such approaches work poorly when the SNR is low and there are few
pilots. The reasons are twofold: firstly, they ignore the fact that
sub-channels share some common scatterers and the associated complex
channel gain, angle, and distance information; secondly, they do not
exploit the structure of VRs. To exploit these, the authors in \cite{Vincent_sparse_modeling,Tang_sparse_modeling}
used a Markov chain model to describe the clustered sparsity of VR
vectors in XL-MIMO ULA systems. Based on the Markov chain model, a
turbo orthogonal approximate message passing (Turbo-OAMP) algorithm
was proposed. By fully exploiting the sparse structure of VR vectors,
the performance of both VR detection and channel estimation was significantly
improved. However, the work in \cite{Tang_sparse_modeling} only considers
a line-of-sight (LoS)-only channel model, and the proposed method
cannot be extended to multipath channels intuitively. Moreover, when
the sensing matrix is not partially orthogonal, the Turbo-OAMP algorithm
will involve a high-dimensional matrix inverse each iteration, which
leads to unacceptable computational overhead.

In this paper, we consider a joint VR detection and channel estimation
problem in a XL-MIMO UPA system at the mmWave band. There are three
main challenges in our considered problem: 1) In the scenario of UPA,
the VR of scatterers will exhibit a 2D clustered sparsity, and thus
a new sparse prior model is needed to capture the 2D clustered sparsity;
2) Since sub-channels corresponding to sub-arrays are correlated via
shared scatterers, the whole channel and the associated channel parameters
should be estimated jointly; 3) Because of the extremely large number
of XL-MIMO, low-complexity algorithm design becomes essential. To
overcome these challenges, we first formulate the considered problem
as a bilinear observation model. Then, we introduce a two-dimensional
(2D) Markov prior model to describe the 2D clustered sparsity of VRs.
Finally, we design an alternating maximum a posteriori (MAP) framework
with acceptable complexity for high-accuracy VR detection and channel
estimation. The main contributions are summarized below.
\begin{itemize}
\item \textbf{Prior design for VRs and the sparse channel: }To exploit the
2D clustered sparsity of VRs, we design a 2D Markov prior model, which
can be treated as an extension of the one-dimensional Markov chain
model in \cite{Vincent_sparse_modeling,Tang_sparse_modeling}. Moreover,
we introduce a hierarchical sparse prior model to capture the sparsity
of the XL-MIMO channel vector in the polar domain. 
\item \textbf{Alternating MAP framework:} We propose an alternating MAP
framework that contains three basic modules: a channel estimation
module, a VR detection module, and a grid update module. The three
modules work alternatively to improve each other's performance. Some
approaches are employed to reduce the complexity of the alternating
MAP. Firstly, inspired by \cite{Duan_IFSBL,Xu_Turbo-IFVBI}, we develop
an inverse-free variational Bayesian inference (IF-VBI) algorithm
as the channel estimation module, where the high-dimensional matrix
inverse operation is avoided via minimizing a relaxed Kullback-Leibler
(KL) divergence. Secondly, in the VR detection module, we use polar-domain
filtering and sub-array grouping methods to achieve matrix dimension
reduction. As such, the alternating MAP framework is able to perform
high-performance VR detection and channel estimation with acceptable
computational overhead.
\item \textbf{Structured expectation propagation (EP) algorithm:} In the
VR detection module, we propose a novel structured EP algorithm based
on the 2D Markov prior. The structured EP algorithm consists of two
modules: a linear minimum-mean-square-error (LMMSE) estimator and
a non-linear MMSE estimator. Compared to the conventional EP \cite{Cespedes_EP},
the proposed structured EP can process more complicated prior information
by using loopy belief propagation in the MMSE estimator.
\end{itemize}
The paper proceeds as follows. In Section II, we show the system model.
In Section III, we focus on the structured prior design for VRs and
the polar-domain sparse channel vector. In Section IV, we present
the novel alternating MAP framework and detail its three basic modules.
Simulations and conclusions are shown in Section V and VI, respectively. 

\textit{Notations:} Lowercase and uppercase bold letterers denote
vectors and matrices, respectively. Let $\left(\cdot\right)^{-1}$,
$\left(\cdot\right)^{T}$, $\left(\cdot\right)^{H}$, $\left\langle \cdot\right\rangle $,
$\left\Vert \cdot\right\Vert $, $\textrm{vec}\left(\cdot\right)$,
and $\textrm{diag}\left(\cdot\right)$ represent the inverse, transpose,
conjugate transpose, expectation, $\ell_{2}\textrm{-norm}$, vectorization,
and diagonalization operations, respectively. $\otimes$ is the Kronecker
product operator and $\odot$ means the Hadamard product operator.
$\Re\left\{ \cdot\right\} $ and $\Im\left\{ \cdot\right\} $ denote
the real and imaginary part of the complex argument, respectively.
$\mathbf{I}_{N}$ is the $N\times N$ dimensional identity matrix
and $\mathbf{1}_{M\times N}$ is the $M\times N$ dimensional all-one
matrix. For a set $\mathbf{\Omega}$ with its cardinal number denoted
by $\left|\mathbf{\Omega}\right|$, $\boldsymbol{x}\triangleq\left[x_{n}\right]_{n\in\mathbf{\Omega}}\in\mathbb{C}^{\left|\mathbf{\Omega}\right|\times1}$
is a vector composed of elements indexed by $\mathbf{\Omega}$. $\mathcal{CN}\left(\boldsymbol{x};\boldsymbol{\mu},\mathbf{\Sigma}\right)$
denotes the complex Gaussian distribution with mean $\boldsymbol{\mu}$
and covariance $\mathbf{\Sigma}$. $\textrm{Ga}\left(x;a,b\right)$
denotes the Gamma distribution with shape parameter $a$ and rate
parameter $b$.
\begin{figure}[t]
\begin{centering}
\includegraphics[width=70mm]{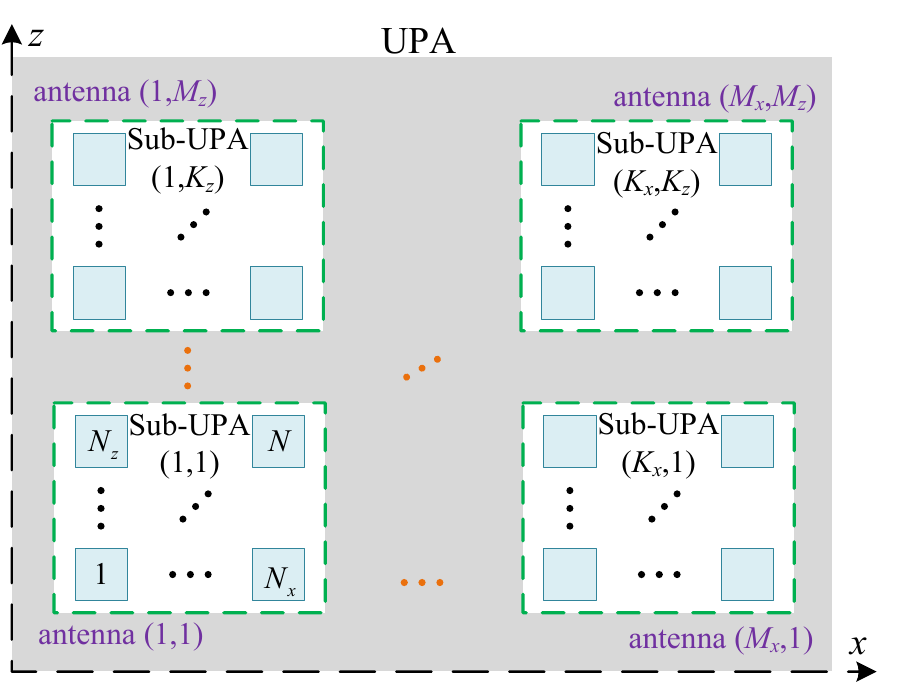}
\par\end{centering}
\caption{\label{fig:Sub_UPAs}The structure of the UPA and the index rule of
antennas and sub-UPAs.}
\end{figure}

\section{System Model}

\subsection{Introduction of the XL-MIMO System}

Consider a XL-MIMO system operating at the mmWave band, where one
base station (BS) serves a single-antenna user,\footnote{The proposed method can be easily extended to the case of multiple
users by assigning orthogonal time/frequency resources to each user.} as shown in Fig. \ref{fig:Illustration_SNS_VR}. The BS is equipped
with a half-wavelength UPA of $M=M_{x}\times M_{z}$ antennas in the
$x\textrm{-}z$ plane. The large-scale UPA is partitioned into $K=K_{x}\times K_{z}$
small-scale sub-UPAs, with each sub-UPA consisting of $N=N_{x}\times N_{z}$
antennas, where $N_{x}=\frac{M_{x}}{K_{x}}$ and $N_{z}=\frac{M_{z}}{K_{z}}$,
as illustrated in Fig. \ref{fig:Sub_UPAs}. The carrier frequency
is $f$ and the wavelength is $\lambda_{c}=\frac{c}{f}$, where $c$
is the speed of light. The spacing between two adjacent antennas is
$d=\frac{\lambda_{c}}{2}$, and the antenna aperture is given by $D=\sqrt{\left(M_{x}-1\right)^{2}+\left(M_{z}-1\right)^{2}}d$.
For convenience, we set the center of the UPA as the origin $\left[0,0,0\right]^{T}$
of the coordinate system. Index the antenna in the bottom left corner
as the $\left(1,1\right)\textrm{-th}$ antenna, as presented in Fig.
\ref{fig:Sub_UPAs}. Define the relative subscript of the $\left(m_{x},m_{z}\right)\textrm{-th}$
antenna as $\left(\delta_{m_{x}},\delta_{m_{z}}\right)=\left(m_{x}-\frac{M_{x}+1}{2},m_{z}-\frac{M_{z}+1}{2}\right)$,
then the coordinates of the $\left(m_{x},m_{z}\right)\textrm{-th}$
antenna can be expressed as $\boldsymbol{p}_{m_{x},m_{z}}=\left[\delta_{m_{x}}d,0,\delta_{m_{z}}d\right]^{T}$.
The user transmits an uplink pilot symbol (assume the pilot is equal
to one without loss of generality), then the received signal at the
BS can be expressed as
\begin{equation}
\boldsymbol{y}=\boldsymbol{h}+\boldsymbol{w},\label{eq:y}
\end{equation}
where $\boldsymbol{h}\in\mathbb{C}^{M\times1}$ is the channel vector
and $\boldsymbol{w}\sim\mathcal{CN}\left(0,1/\gamma\mathbf{I}_{M}\right)$
is the additive white Gaussian noise (AWGN) with variance $1/\gamma$.
As near-field propagation and spatial non-stationary effects exist,
the structure of the channel vector is quite different from that of
far-field massive MIMO channels and will be exploited to significantly
improve the channel estimation performance.
\begin{figure}[t]
\begin{centering}
\includegraphics[width=70mm]{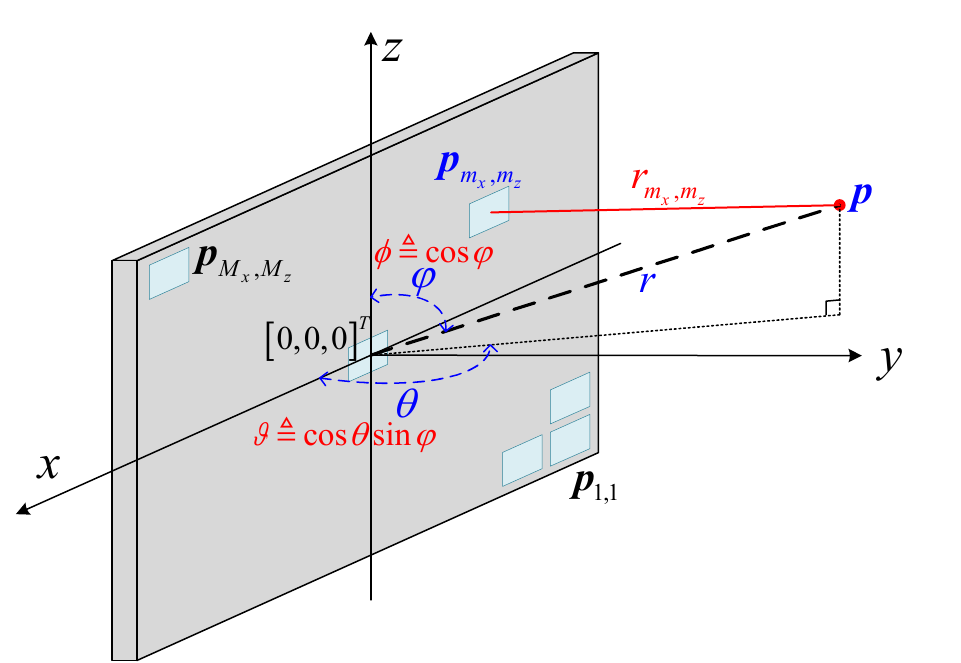}
\par\end{centering}
\caption{\label{fig:Coordinate_system} The distance, azimuth angle, and elevation
angle of a scatterer in the coordinate system.}
\end{figure}

\subsection{Near-field Spatial Non-stationary Channel Model}

In XL-MIMO systems at the mmWave band, the Rayleigh distance is hundreds
of meters \cite{Liu_NF_Tutorial}, and thus near-field propagation
cannot be neglected. The impact of near-field propagation is primarily
reflected in the steering vector. Let $\theta$, $\varphi$, and $r$
denote the azimuth angle, elevation angle, and distance of a scatterer
with respect to the origin, respectively, as illustrated in Fig. \ref{fig:Coordinate_system}.
The coordinates of the scatterer can be expressed as $\boldsymbol{p}=\left[r\cos\theta\sin\varphi,r\sin\theta\sin\varphi,r\cos\varphi\right]^{T}$.
Then the distance between the scatterer and the $\left(m_{x},m_{z}\right)\textrm{-th}$
antenna of the UPA is given by
\begin{align}
 & r_{m_{x},m_{z}}=\left\Vert \boldsymbol{p}-\boldsymbol{p}_{m_{x},m_{z}}\right\Vert =\label{eq:r_m_n_original}\\
 & \sqrt{r^{2}+\delta_{m_{x}}^{2}d^{2}+\delta_{m_{z}}^{2}d^{2}-2r\delta_{m_{x}}d\cos\theta\sin\varphi-2r\delta_{m_{z}}d\cos\varphi}.\nonumber 
\end{align}
 Referring to the uniform spherical wave (USW) model \cite{Starer_USW1,Liu_NF_Tutorial},
the steering vector is derived as
\begin{align}
\boldsymbol{a}\left(\theta,\varphi,r\right)= & \frac{1}{\sqrt{M}}\left[e^{-j\frac{2\pi}{\lambda_{c}}\left(r_{1,1}-r\right)},\ldots,e^{-j\frac{2\pi}{\lambda_{c}}\left(r_{M_{x},1}-r\right)},\ldots,\right.\nonumber \\
 & \left.e^{-j\frac{2\pi}{\lambda_{c}}\left(r_{1,M_{z}}-r\right)},\ldots,e^{-j\frac{2\pi}{\lambda_{c}}\left(r_{M_{x},M_{z}}-r\right)}\right]^{T}.\label{eq:Ar_original}
\end{align}
The Fresnel approximation is usually introduced to simplify the complicated
expressions in (\ref{eq:r_m_n_original}) and (\ref{eq:Ar_original}).
Based on the Fresnel approximation \cite{Selvan_Fresnel_distance},
the distance $r_{m_{x},m_{z}}$ is approximate to
\begin{align}
r_{m_{x},m_{z}}\approx & r-\delta_{m_{x}}d\cos\theta\sin\varphi+\frac{\delta_{m_{x}}^{2}d^{2}\left(1-\cos^{2}\theta\sin^{2}\varphi\right)}{2r}\nonumber \\
 & -\delta_{m_{z}}d\cos\varphi+\frac{\delta_{m_{z}}^{2}d^{2}\left(1-\cos^{2}\varphi\right)}{2r}.
\end{align}
Let $\vartheta\triangleq\cos\theta\sin\varphi$ and $\phi\triangleq\cos\varphi$,
the steering vector under the Fresnel approximation can be obtained
as
\begin{equation}
\boldsymbol{a}\left(\vartheta,\phi,r\right)=\frac{1}{\sqrt{M}}\boldsymbol{a}_{z}\left(\phi,r\right)\otimes\boldsymbol{a}_{x}\left(\vartheta,r\right),
\end{equation}
with
\begin{subequations}
\begin{align}
\left[\boldsymbol{a}_{x}\left(\vartheta,r\right)\right]_{m_{x}} & =e^{-j\frac{2\pi}{\lambda_{c}}\left(-\delta_{m_{x}}d\vartheta+\frac{\delta_{m_{x}}^{2}d^{2}\left(1-\vartheta^{2}\right)}{2r}\right)},\\
\left[\boldsymbol{a}_{z}\left(\phi,r\right)\right]_{m_{z}} & =e^{-j\frac{2\pi}{\lambda_{c}}\left(-\delta_{m_{z}}d\phi+\frac{\delta_{m_{z}}^{2}d^{2}\left(1-\phi^{2}\right)}{2r}\right)},
\end{align}
\end{subequations}
where $\left[\boldsymbol{a}_{x}\left(\vartheta,r\right)\right]_{m_{x}}$
is the $m_{x}\textrm{-th}$ element of $\boldsymbol{a}_{x}\left(\vartheta,r\right)$
for $m_{x}=1,\ldots,M_{x}$, and $\left[\boldsymbol{a}_{z}\left(\phi,r\right)\right]_{m_{z}}$
is the $m_{z}\textrm{-th}$ element of $\boldsymbol{a}_{z}\left(\phi,r\right)$
for $m_{z}=1,\ldots,M_{z}$. The work in \cite{Selvan_Fresnel_distance}
has verified that the Fresnel approximation is accurate enough when
the distance between the scatterer and the BS is large than the Fresnel
distance. Note that the Fresnel distance $0.5\sqrt{\frac{D^{3}}{\lambda_{c}}}$
is much smaller than the Rayleigh distance $\frac{2D^{2}}{\lambda_{c}}$.
For instance, if the antenna aperture is $D=1\ \textrm{m}$ and the
carrier frequency is $f=30\ \textrm{GHz}$, then the Rayleigh distance
will be $200\ \textrm{m}$, while the Fresnel distance will be only
$5\ \textrm{m}$ that can nearly be neglected.

Besides, the spatial non-stationary effect is also obvious in XL-MIMO
systems \cite{Carvalho_NSN}, i.e., different regions of the UPA will
receive different levels of power due to different scatterers they
can see. Assume there are $L$ paths between the user and the BS,
and we only focus on the last-hop scatterer of the paths. We define
a binary matrix
\begin{equation}
\mathbf{V}_{l}=\left[\begin{array}{cccc}
v_{l,1,1} & v_{l,1,2} & \cdots & v_{l,1,K_{z}}\\
v_{l,2,1} & v_{l,2,2} & \cdots & v_{l,2,K_{z}}\\
\vdots & \vdots & \ddots & \vdots\\
v_{l,K_{x},1} & v_{l,K_{x},2} & \cdots & v_{l,K_{x},K_{z}}
\end{array}\right],\label{eq:V_l}
\end{equation}
 to represent the visibility region of scatterer $l$ for $l=1,\ldots,L$,
where $v_{l,k_{x},k_{z}}=1$ indicates scatterer $l$ can see sub-UPA
$\left(k_{x},k_{z}\right)$ and $v_{l,k_{x},k_{z}}=0$ indicates the
opposite.

Based on these, the channel vector $\boldsymbol{h}$ can be modeled
as
\begin{equation}
\boldsymbol{h}=\sum_{l=1}^{L}x_{l}\boldsymbol{a}\left(\vartheta_{l},\phi_{l},r_{l}\right)\odot\boldsymbol{u}_{l},\label{eq:h}
\end{equation}
where $x_{l}$ denotes the complex gain of the $l\textrm{-th}$ path
and $r_{l}$ is the distance of scatterer $l$. $\vartheta_{l}\triangleq\cos\theta_{l}\sin\varphi_{l}$
and $\phi_{l}\triangleq\cos\varphi_{l}$, in which $\theta_{l}$ and
$\varphi_{l}$ are the azimuth and elevation angles of scatterer $l$,
respectively. $\boldsymbol{a}\left(\vartheta_{l},\phi_{l},r_{l}\right)$
is the steering vector simulated by scatterer $l$ and $\boldsymbol{u}_{l}\triangleq\textrm{vec}\left(\mathbf{V}_{l}\otimes\mathbf{1}_{N_{x}\times N_{z}}\right)$
is the element-level VR vector. 

Moreover, there is spatial correlation among adjacent sub-UPAs, i.e.,
adjacent sub-UPAs may be visible to the same scatterer with a high
probability. In this case, the binary matrix $\mathbf{V}_{l},\forall l$
exhibits a 2D clustered sparsity. Therefore, it is essential to design
a sparse prior model to capture such a structured sparsity.

\section{Prior Design for VRs and the Polar-domain Channel}

In this section, we first introduce a polar-domain sparse representation
method for the XL-MIMO channel. Then, we present sparse probability
models for both VRs and the sparse channel vector. Finally, the joint
VR detection and channel estimation problem is formulated as a MAP
estimation problem.

\subsection{Polar-domain Sparse Representation Method}

We adopt the grid-based method to obtain a sparse representation of
the XL-MIMO channel for high-accuracy channel estimation. In \cite{Cui_polar_grid,Yang_3D_polar},
the authors verified that the near-field channel exhibits sparsity
in the polar domain, which accounts for both angle and distance information
simultaneously. Inspired by this, we design a 3-D polar-domain grid
for the case of UPA. Specifically, we first introduce a angle gird
of $M_{1}\times M_{2}$ angle points, in which the sampling points
$\left\{ \bar{\vartheta}_{m_{1}}\right\} _{m_{1}=1}^{M_{1}}$ and
$\left\{ \bar{\phi}_{m_{2}}\right\} _{m_{2=1}}^{M_{2}}$ are uniformly
distributed within $[-1,1]$. Then, at the sampled angle $\left(\bar{\vartheta}_{m_{1}},\bar{\phi}_{m_{2}}\right),\forall m_{1},m_{2}$,
the distance sampling points $\left\{ \bar{r}_{m_{1},m_{2},n}\right\} _{n=1}^{N_{m_{1},m_{2}}}$
can be obtained using Algorithm 1 in \cite{Cui_polar_grid}, where
$N_{m_{1},m_{2}}$ denotes the number of sampled distances at direction
$\left(\bar{\vartheta}_{m_{1}},\bar{\phi}_{m_{2}}\right)$. Based
on these, the polar-domain grid points can be expressed as
\begin{equation}
\begin{aligned}\bar{\boldsymbol{\Xi}}\triangleq & \Bigl[\bigl[\bar{\vartheta}_{1},\bar{\phi}_{1},\bar{r}_{1}\bigr];\ldots;\bigl[\bar{\vartheta}_{1},\bar{\phi}_{1},\bar{r}_{N_{1,1}}\bigr];\ldots\\
 & \bigl[\bar{\vartheta}_{m_{1}},\bar{\phi}_{m_{2}},\bar{r}_{1}\bigr];\ldots;\bigl[\bar{\vartheta}_{m_{1}},\bar{\phi}_{m_{2}},\bar{r}_{N_{m_{1},m_{2}}}\bigr];\ldots\\
 & \bigl[\bar{\vartheta}_{M_{1}},\bar{\phi}_{M_{2}},\bar{r}_{1}\bigr];\ldots;\bigl[\bar{\vartheta}_{M_{1}},\bar{\phi}_{M_{2}},\bar{r}_{N_{M_{1},M_{2}}}\bigr]\Bigr].
\end{aligned}
\label{eq:Polar-grid}
\end{equation}
The total number of grid points is given by $Q=\sum_{m_{1}=1}^{M_{1}}\sum_{m_{2}=1}^{M_{2}}N_{m_{1},m_{2}}$.
To simply the notation, we use $q$ to index the grid points in $\bar{\boldsymbol{\Xi}}$,
and the $q\textrm{-th}$ polar-domain grid point (i.e., the $q\textrm{-th}$
row of $\bar{\boldsymbol{\Xi}}$) is denoted by $\left[\bar{\vartheta}_{q},\bar{\phi}_{q},\bar{r}_{q}\right]$
for $q=1,\ldots,Q$.

However, in practice, the true angle-distances of scatterers usually
do not lie exactly on the discrete polar-domain grid points. As a
result, there exists a mismatch between the true angle-distance and
its nearest grid point, which will lead to an energy leakage effect.
In the high SNR regions, the leakage effect caused by angle and distance
mismatch is especially obvious, which cannot be neglected compared
to the noise power. Therefore, we introduce a dynamic polar-domain
grid, denoted by $\boldsymbol{\Xi}\triangleq\left[\boldsymbol{\vartheta},\boldsymbol{\phi},\boldsymbol{r}\right]$,
instead of only using a fixed grid, where $\boldsymbol{\vartheta}\triangleq\left[\vartheta_{1},\ldots,\vartheta_{Q}\right]^{T}$,
$\boldsymbol{\phi}\triangleq\left[\phi_{1},\ldots,\phi_{Q}\right]^{T}$,
and $\boldsymbol{r}\triangleq\left[r_{1},\ldots,r_{Q}\right]^{T}$
denote angle and distance dynamic grid vectors. The fixed grid $\bar{\boldsymbol{\Xi}}$
is chosen as the initial value of $\boldsymbol{\Xi}$, and the grid
parameters are updated via gradient ascent during algorithm design. 

With the definition of the dynamic polar-domain grid, we can obtain
a polar-domain sparse basis as
\begin{equation}
\left[\mathbf{A}\left(\boldsymbol{\Xi}\right)\right]_{q}=\left[\boldsymbol{a}\left(\vartheta_{q},\phi_{q},r_{q}\right)\right],\forall q,
\end{equation}
where $\left[\mathbf{A}\left(\boldsymbol{\Xi}\right)\right]_{q}$
denotes the $q\textrm{-th}$ column of $\mathbf{A}\left(\boldsymbol{\Xi}\right)\in\mathbb{C}^{M\times Q}$.
Besides, let $\mathbf{V}_{q}\in\left\{ 0,1\right\} ^{K_{x}\times K_{z}}$
represent the VR of the scatterer lying around the $q\textrm{-th}$
grid point and denote the collection of VRs as $\mathbf{V}\triangleq\left\{ \mathbf{V}_{q}\mid\forall q\right\} $.
Define $\mathbf{U}\left(\mathbf{V}\right)\triangleq\left[\boldsymbol{u}_{1},\ldots,\boldsymbol{u}_{Q}\right]\in\mathbb{C}^{M\times Q}$
as the VR dictionary corresponding to the polar-domain grid, where
$\boldsymbol{u}_{q}\triangleq\textrm{vec}\left(\mathbf{V}_{q}\otimes\mathbf{1}_{N_{x}\times N_{z}}\right),\forall q$.
Then the sparse representation of the channel vector in (\ref{eq:h})
can be obtained as
\begin{align}
\boldsymbol{h} & =\left[\mathbf{A}\left(\boldsymbol{\Xi}\right)\odot\mathbf{U}\left(\mathbf{V}\right)\right]\boldsymbol{x},\label{eq:h_sparse}
\end{align}
where $\boldsymbol{x}\in\mathbb{C}^{Q\times1}$ is the polar-domain
sparse channel vector, which has only $L\ll Q$ non-zero elements
corresponding to $L$ paths. Specifically, the $q\textrm{-th}$ element
of $\boldsymbol{x}$, denoted by $x_{q}$, is the complex gain of
the channel path with the corresponding scatterer lying around the
$q\textrm{-th}$ grid point. 

Note that the transform matrix in (\ref{eq:h_sparse}), denoted by
$\mathbf{A}\left(\boldsymbol{\Xi}\right)\odot\mathbf{U}\left(\mathbf{V}\right)$,
is quite different from that in \cite{Cui_polar_grid,Yang_3D_polar}
due to the existing of spatial non-stationary. The VR dictionary $\mathbf{U}\left(\mathbf{V}\right)$
provides the VR information of each scatterer lying in the polar-domain
grid.

\subsection{Sparse Probability Model}

In this subsection, we first design a 2D Markov model to capture the
2D clustered sparsity of VRs. Besides, a hierarchical sparse prior
model is used to describe the sparsity of the polar-domain channel
vector. Finally, we obtain the joint distribution of all variables.

\subsubsection{2D Markov model for VRs }

Since the VR of scatterers exhibits a 2D clustered sparsity, the non-zero
elements in $\mathbf{V}_{q},\forall q$ will concentrate on a few
clusters. This implies that $v_{q,k_{x},k_{z}}$ depends on both $v_{q,k_{x}-1,k_{z}}$
and $v_{q,k_{x},k_{z}-1}$. Specifically, if $v_{q,k_{x},k_{z}-1}=1$
or $v_{q,k_{x},k_{z}-1}=1$, there is a higher probability that $v_{q,k_{x},k_{z}}=1$.
Such a structured sparsity can be modeled using the 2D Markov model
\cite{Fornasini_2D_MM},
\begin{align}
p\left(\mathbf{V}_{q}\right)= & p\left(v_{q,1,1}\right)\prod_{k_{x}=2}^{K_{x}}\prod_{k_{z}=1}^{K_{z}}p\left(v_{q,k_{x},k_{z}}\mid v_{q,k_{x}-1,k_{z}}\right)\nonumber \\
 & \times\prod_{k_{x}=1}^{K_{x}}\prod_{k_{z}=2}^{K_{z}}p\left(v_{q,k_{x},k_{z}}\mid v_{q,k_{x},k_{z}-1}\right),\forall q,\label{eq:2D-MM}
\end{align}
with the transition probability given by
\begin{subequations}
\begin{align}
p_{01}^{x} & =p\left(v_{q,k_{x},k_{z}}=1\mid v_{q,k_{x}-1,k_{z}}=0\right),\\
p_{10}^{x} & =p\left(v_{q,k_{x},k_{z}}=0\mid v_{q,k_{x}-1,k_{z}}=1\right),\\
p_{01}^{z} & =p\left(v_{q,k_{x},k_{z}}=1\mid v_{q,k_{x},k_{z}-1}=0\right),\\
p_{10}^{z} & =p\left(v_{q,k_{x},k_{z}}=0\mid v_{q,k_{x},k_{z}-1}=1\right).
\end{align}
\end{subequations}
Note that \cite{Fornasini_2D_MM} has verified that a set of $\left\{ p_{01}^{x},p_{10}^{x},p_{01}^{z},p_{10}^{z}\right\} $
can be found to satisfy the steady-state condition of the 2D Markov
model, i.e., $p\left(v_{q,k_{x},k_{z}}=1\right)=\kappa,\forall k_{x},k_{z}$,
where $\kappa$ defined as the visibility probability shows the sparsity
level of $\mathbf{V}_{q}$. As such, the initial distribution $p\left(v_{q,1,1}\right)$
is set to be the steady-state distribution, $p\left(v_{q,1,1}=1\right)=\kappa$.
The factor graph of the 2D Markov model is presented in Fig. \ref{fig:Factor-graph-Module-B}.

The value of $\left\{ p_{01}^{x},p_{10}^{x},p_{01}^{z},p_{10}^{z}\right\} $
will affect the structure of clusters. Specifically, smaller $p_{10}^{x}$
and $p_{10}^{z}$ imply a larger average cluster size, and smaller
$p_{01}^{x}$ and $p_{01}^{z}$ imply a larger average gap between
two adjacent clusters. Therefore, the 2D Markov model has the flexibility
to characterize the 2D clustered sparsity of $\mathbf{V}_{q}$.

\subsubsection{Hierarchical sparse prior for the channel}

\begin{figure}[t]
\begin{centering}
\includegraphics[width=80mm]{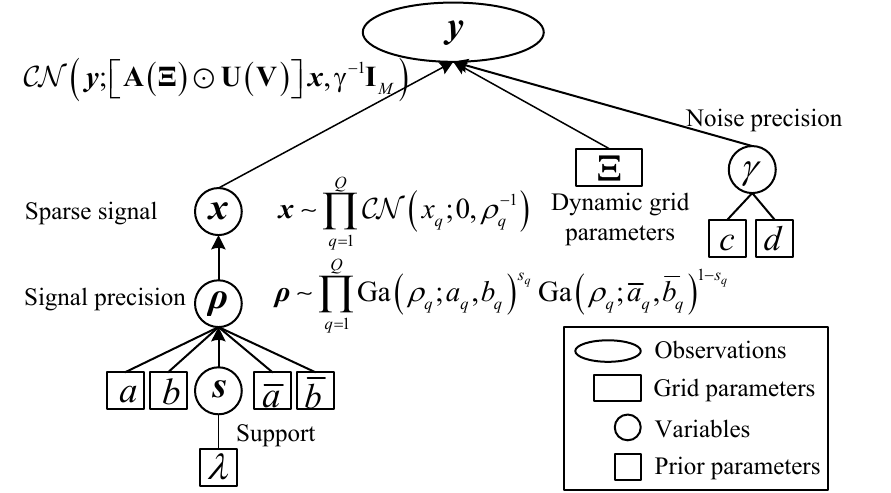}
\par\end{centering}
\caption{\label{fig:Sparse_prior_model}Illustration of the hierarchical sparse
prior model.}
\end{figure}

We introduce a hierarchical sparse prior model to capture the sparsity
of the polar-domain channel vector $\boldsymbol{x}$, as illustrated
in Fig. \ref{fig:Sparse_prior_model}. Specifically, let $\boldsymbol{s}\triangleq\left[s_{1},\ldots,s_{Q}\right]^{T}$
denote the support of $\boldsymbol{x}$, where $s_{q}=1$ indicates
$x_{q}$ is non-zero and $s_{q}=0$ indicates the opposite. Let $\boldsymbol{\rho}\triangleq\left[\rho_{1},\ldots,\rho_{Q}\right]^{T}$
denote the precision vector of $\boldsymbol{x}$, where $1/\rho_{q}$
gives the variance of $x_{q}$. The variables $\boldsymbol{x}$, $\boldsymbol{\rho}$,
and $\boldsymbol{s}$ form a Markov chain, denoted by $\boldsymbol{s}\rightarrow\boldsymbol{\rho}\rightarrow\boldsymbol{x}$,
and the joint distribution can be expressed as
\begin{equation}
p\left(\boldsymbol{x},\boldsymbol{\rho},\boldsymbol{s}\right)=\underbrace{p\left(\boldsymbol{s}\right)}_{\textrm{Support}}\underbrace{p\left(\boldsymbol{\rho}\mid\boldsymbol{s}\right)}_{\textrm{Precision}}\underbrace{p\left(\boldsymbol{x}\mid\boldsymbol{\rho}\right)}_{\textrm{Sparse\ signal}}.\label{eq:p(x,rou,s)}
\end{equation}
For an independent sparse structure, a Bernoulli prior is usually
used to model the support vector \cite{Yuan_BGprior1,Yuan_BGprior2},
\begin{equation}
p\left(\boldsymbol{s}\right)=\prod_{q=1}^{Q}\left(\lambda_{q}\right)^{s_{q}}\left(1-\lambda_{q}\right)^{1-s_{q}},
\end{equation}
where $\lambda_{q}$ gives the probability that $s_{q}=1$.

The conditional distribution $p\left(\boldsymbol{\rho}\mid\boldsymbol{s}\right)$
is given by
\begin{equation}
p\left(\boldsymbol{\rho}\mid\boldsymbol{s}\right)=\prod_{q=1}^{Q}\textrm{Ga}\left(\rho_{q};a_{q},b_{q}\right)^{s_{q}}\textrm{Ga}\left(\rho_{q};\overline{a}_{q},\overline{b}_{q}\right)^{1-s_{q}}.
\end{equation}
We use two different Gamma distributions to model $\rho_{q}$ according
to the value of $s_{q}$. When $s_{q}=1$, the element $x_{q}$ is
non-zero. In this case, the corresponding parameters of the precision
$\rho_{q}$, denoted by $a_{q}$ and $b_{q}$, should be chosen to
satisfy $\frac{a_{q}}{b_{q}}=\mathbb{E}\left[\rho_{q}\right]=\mathcal{O}\left(1\right)$
such that the variance of $x_{q}$ is also $\mathcal{O}\left(1\right)$.
When $s_{q}=0$, the element $x_{q}$ is zero or close to zero. In
this case, the corresponding parameters of the precision $\rho_{q}$,
denoted by $\overline{a}_{q}$ and $\overline{b}_{q}$, should be
chosen to satisfy $\frac{\overline{a}_{q}}{\overline{b}_{q}}=\mathbb{E}\left[\rho_{q}\right]\gg1$
such that the variance of $x_{q}$ is close to zero. The motivation
for choosing Gamma is that the Gamma distribution is a conjugate of
the Gaussian prior, which facilitates closed-form Bayesian inference
\cite{Tipping_SBL,Ji_SBL,Tzikas_VBI}.

The conditional distribution $p\left(\boldsymbol{x}\mid\boldsymbol{\rho}\right)$
is given by
\begin{equation}
p\left(\boldsymbol{x}\mid\boldsymbol{\rho}\right)=\prod_{q=1}^{Q}p\left(x_{q}\mid\rho_{q}\right)=\prod_{q=1}^{Q}\mathcal{CN}\left(x_{q};0,\rho_{q}^{-1}\right).
\end{equation}
 In practice, the exact distribution of each element of $\boldsymbol{x}$
is usually unknown. Nevertheless, it is reasonable to choose a hierarchical
Gaussian prior with Gamma-distributed precision for $\boldsymbol{x}$.
Firstly, the Gaussian prior facilitates low-complexity algorithm design.
Secondly, the existing VBI-type algorithms derived from such a hierarchical
prior are well known to be insensitive to the true distribution of
$\boldsymbol{x}$ \cite{Tipping_SBL,Ji_SBL,Tzikas_VBI}.

Moreover, we use a Gamma distribution with parameters $c$ and $d$
to model the noise precision \cite{Tipping_SBL,Ji_SBL,Tzikas_VBI},
\begin{equation}
p\left(\gamma\right)=\textrm{Ga}\left(\gamma;c,d\right).\label{eq:p(gamma)}
\end{equation}
According to the existing works \cite{LiuAn_CE_Turbo_VBI,LiuAn_directloc_vehicles,Xu_Turbo-IFVBI,Xu_SLA_VBI},
the hierarchical sparse prior model is robust w.r.t. the the imperfect
prior information in practice, and it is tractable to enable low-complexity
and high-performance algorithm design.

\subsubsection{The joint distribution}

The joint distribution of the random variables discussed above is
given by
\begin{align}
p\left(\boldsymbol{y},\mathbf{V},\boldsymbol{x},\boldsymbol{\rho},\boldsymbol{s},\gamma;\boldsymbol{\Xi}\right)= & p\left(\boldsymbol{y}\mid\mathbf{V},\boldsymbol{x},\gamma;\boldsymbol{\Xi}\right)\nonumber \\
 & \times\prod_{q=1}^{Q}p\left(\mathbf{V}_{q}\right)p\left(\boldsymbol{x},\boldsymbol{\rho},\boldsymbol{s}\right)p\left(\gamma\right),\label{eq:joint distribution}
\end{align}
with the likelihood function given by
\begin{align*}
p\left(\boldsymbol{y}\mid\mathbf{V},\boldsymbol{x},\gamma;\boldsymbol{\Xi}\right)= & \mathcal{CN}\left(\boldsymbol{y};\left[\mathbf{A}\left(\boldsymbol{\Xi}\right)\varodot\mathbf{U}\left(\mathbf{V}\right)\right]\boldsymbol{x},\gamma^{-1}\mathbf{I}_{M}\right).
\end{align*}

\subsection{The MAP Problem Formulation}

Using the polar-domain sparse representation in (\ref{eq:h_sparse}),
the received signal in (\ref{eq:y}) can be written into an observation
model with unknown VRs and uncertain grid parameters in the sensing
matrix,
\begin{equation}
\boldsymbol{y}=\left[\mathbf{A}\left(\boldsymbol{\Xi}\right)\odot\mathbf{U}\left(\mathbf{V}\right)\right]\boldsymbol{x}+\boldsymbol{w}.\label{eq:y_MAP}
\end{equation}
Since it is quite difficult to perform Bayesian inference for all
variables, we aim at computing the MAP estimate of the channel vector
$\boldsymbol{x}$, the VRs $\mathbf{V}$, and the grid parameters
$\boldsymbol{\Xi}$ given the observation $\boldsymbol{y}$, i.e.,
\begin{equation}
\boldsymbol{x}^{\dagger},\mathbf{V}^{\dagger},\boldsymbol{\Xi}^{\dagger}=\underset{\boldsymbol{x},\mathbf{V},\boldsymbol{\Xi}}{\arg\max}\ln p\left(\boldsymbol{y},\mathbf{V},\boldsymbol{x},\boldsymbol{\rho},\boldsymbol{s},\gamma;\boldsymbol{\Xi}\right),\label{eq:MAP_problem}
\end{equation}
where the joint distribution $p\left(\boldsymbol{y},\mathbf{V},\boldsymbol{x},\boldsymbol{\rho},\boldsymbol{s},\gamma;\boldsymbol{\Xi}\right)$
is given in (\ref{eq:joint distribution}). However, it is still intractable
to directly solve the MAP problem in (\ref{eq:MAP_problem}) since
different variables have quite different priors and likelihood functions.
To address this problem, we update the MAP estimate of each variable
in an alternating way. Specifically, for given $\hat{\mathbf{V}}$
and $\hat{\boldsymbol{\Xi}}$, we perform Bayesian inference to compute
the marginal posterior distribution of $\boldsymbol{x}$ and $\gamma$,
denoted by $q\left(\boldsymbol{x}\mid\boldsymbol{y},\hat{\mathbf{V}};\hat{\boldsymbol{\Xi}}\right)$
and $q\left(\gamma\mid\boldsymbol{y},\hat{\mathbf{V}};\hat{\boldsymbol{\Xi}}\right)$,
respectively. Then the MAP estimate of $\boldsymbol{x}$ is obtained
as $\hat{\boldsymbol{x}}=\textrm{arg}\max_{\boldsymbol{x}}q\left(\boldsymbol{x}\mid\boldsymbol{y},\hat{\mathbf{V}};\hat{\boldsymbol{\Xi}}\right)$
and the MMSE estimate\footnote{We find that adopting the MMSE estimator for the noise precision makes
the algorithm converge faster. However, we still call the proposed
algorithm ``alternating MAP'' since most variables are updated based
on MAP.} of $\gamma$ is obtained as $\hat{\gamma}=\int\gamma q\left(\gamma\mid\boldsymbol{y},\hat{\mathbf{V}};\hat{\boldsymbol{\Xi}}\right)\textrm{d}\gamma$.
For given $\hat{\boldsymbol{x}}$, $\hat{\gamma}$, and $\hat{\boldsymbol{\Xi}}$,
we compute the approximate posterior distribution $p\left(\mathbf{V}_{q}\mid\boldsymbol{y},\hat{\boldsymbol{x}},\hat{\gamma};\hat{\boldsymbol{\Xi}}\right),\forall q$
using a structured EP algorithm and obtain the MAP estimate of $\mathbf{V}_{q}$
as $\mathbf{\hat{V}}_{q}=\textrm{arg}\max_{\mathbf{V}_{q}}p\left(\mathbf{V}_{q}\mid\boldsymbol{y},\hat{\boldsymbol{x}},\hat{\gamma};\hat{\boldsymbol{\Xi}}\right)$.
Meanwhile, based on $\hat{\boldsymbol{x}}$, $\hat{\gamma}$, and
$\hat{\mathbf{V}}$, the grid parameters are obtained by the MAP estimator
as $\hat{\boldsymbol{\Xi}}=\textrm{arg}\max_{\boldsymbol{\Xi}}\ln p\left(\boldsymbol{\Xi}\mid\boldsymbol{y},\hat{\mathbf{V}},\hat{\boldsymbol{x}},\hat{\gamma}\right)$.

There are two main challenges during algorithm design for the MAP
problem. Firstly, the conventional Bayesian inference algorithms (when
updating $\boldsymbol{x}$ and $\gamma$) usually involve a matrix
inverse each iteration, whose complexity is unacceptable when the
number of XL-MIMO array is large. Secondly, the conventional EP algorithm
is unable to compute the posterior distribution of $\mathbf{V}$ since
the 2D Markov prior is quite complicated and the corresponding factor
graph contains loops. Therefore, it is exceedingly challenging to
design a high-efficient algorithm with acceptable complexity for the
considered MAP problem. In the following section, we propose a novel
alternating MAP framework to overcome these challenges.

\section{The Alternating MAP Framework}

\subsection{Outline of the Alternating MAP Framework}

\begin{figure}[t]
\begin{centering}
\includegraphics[width=80mm]{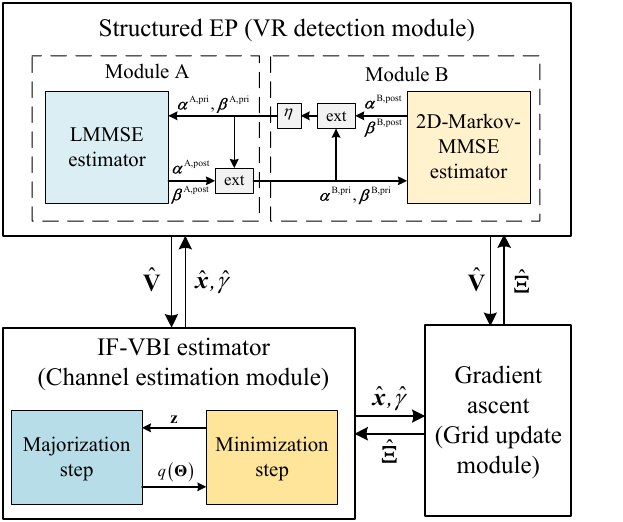}
\par\end{centering}
\caption{\label{fig:The-alternating-MAP}The alternating MAP framework and
its three basic modules.}
\end{figure}

The alternating MAP framework consists of three basic modules: the
channel estimation module, the VR detection module, and the grid update
module, as illustrated in Fig. \ref{fig:The-alternating-MAP}. The
three modules work alternatively until convergence. In the following,
we give a brief introduction to the three modules.

\begin{itemize}
\item \textbf{Channel estimation module:} It is a low-complexity IF-VBI
estimator that avoids the matrix inverse via minimizing a relaxed
KL divergence. For given $\hat{\mathbf{V}}$ and $\hat{\boldsymbol{\Xi}}$,
the IF-VBI estimator optimizes the marginal posterior distribution
of $\boldsymbol{x}$, $\boldsymbol{\rho}$, $\boldsymbol{s}$, and
$\gamma$ alternatively. Then, the MAP estimate of $\boldsymbol{x}$
and the MMSE estimate of $\gamma$ are obtained based on the corresponding
marginal posterior distribution.
\item \textbf{VR detection module:} It is a structured EP algorithm that
combines the LMMSE estimator and the 2D-Markov-MMSE estimator via
a turbo approach. For given $\hat{\boldsymbol{x}}$, $\hat{\gamma}$,
and $\hat{\boldsymbol{\Xi}}$, the structured EP computes the posterior
distribution of $\mathbf{V}_{q},\forall q$ approximately by combining
observation information and 2D Markov prior information. And $\hat{\mathbf{V}}$
is updated by maximizing the approximate posterior distribution.
\item \textbf{Grid update module:} Based on the $\hat{\boldsymbol{x}}$,
$\hat{\gamma}$, and $\hat{\mathbf{V}}$ obtained in the other two
modules, compute the gradient of $\ln p\left(\boldsymbol{y}\mid\hat{\mathbf{V}},\hat{\boldsymbol{x}},\hat{\gamma},;\boldsymbol{\Xi}\right)$
w.r.t. $\boldsymbol{\Xi}$, then update $\hat{\boldsymbol{\Xi}}$
via gradient ascent method.
\end{itemize}

\subsection{IF-VBI Estimator (Channel Estimation Module)}

In this module, the transform matrix in (\ref{eq:y_MAP}) is fixed
since $\hat{\mathbf{V}}$ and $\hat{\boldsymbol{\Xi}}$ are given.
We define $\mathbf{F}\left(\hat{\mathbf{V}},\hat{\boldsymbol{\Xi}}\right)\triangleq\mathbf{A}\left(\hat{\mathbf{V}}\right)\odot\mathbf{U}\left(\hat{\boldsymbol{\Xi}}\right)$
and omit $\hat{\mathbf{V}},\hat{\boldsymbol{\Xi}}$ in $\mathbf{F}\left(\hat{\mathbf{V}},\hat{\boldsymbol{\Xi}}\right)$
for simplicity. Now the observation model in (\ref{eq:y_MAP}) can
be reformulated as $\boldsymbol{y}=\mathbf{F}\boldsymbol{x}+\boldsymbol{w}$,
and the corresponding channel estimation problem is a standard compressive
sensing (CS) problem. Many methods have been proposed to solve such
a CS problem, among which the Turbo-OAMP in \cite{Vincent_sparse_modeling,Tang_sparse_modeling}
and AMP in \cite{Bayati_AMP} can achieve the best trade-off between
the performance and complexity. However, since the sensing matrix
$\mathbf{F}$ is neither partially orthogonal nor i.i.d., both Turbo-OAMP
and AMP involve a high-dimensional matrix inverse each iteration,
which leads to unacceptable computational overhead for XL-MIMO. To
overcome this issue, we design a low-complexity IF-VBI algorithm.

For convenience, we define $\mathbf{\Theta}\triangleq\left\{ \boldsymbol{x},\boldsymbol{\rho},\boldsymbol{s},\gamma\right\} $
as the collection of hidden variables. Let $\mathbf{\Theta}^{k}$
denote an individual variable in $\mathbf{\Theta}$ and let $\boldsymbol{\mathcal{H}}\triangleq\left\{ k\mid\forall\mathbf{\Theta}^{k}\in\mathbf{\Theta}\right\} $.
The mead-filed VBI method uses the variational distribution $q\left(\mathbf{\Theta}\right)$
to approximate the true posterior distribution $p\left(\mathbf{\Theta}\mid\boldsymbol{y}\right)$.
And $q\left(\mathbf{\Theta}\right)$ is optimized via minimizing the
KL divergence between $q\left(\mathbf{\Theta}\right)$ and $p\left(\mathbf{\Theta}\mid\boldsymbol{y}\right)$
under a factorized form constraint as
\begin{align}
\min_{q\left(\mathbf{\Theta}\right)}\mathrm{KL}\left(q\parallel p\right) & \triangleq\int q\left(\mathbf{\Theta}\right)\ln\frac{q\left(\mathbf{\Theta}\right)}{p\left(\mathbf{\Theta}\mid\boldsymbol{y}\right)}\mathrm{d}\boldsymbol{v},\nonumber \\
s.t.\hspace{8mm}q\left(\mathbf{\Theta}\right) & =\prod_{k\in\boldsymbol{\mathcal{H}}}q\left(\mathbf{\Theta}^{k}\right),\label{eq:Problem_VBI}
\end{align}
where $q\left(\mathbf{\Theta}\right)=\prod_{k\in\boldsymbol{\mathcal{H}}}q\left(\mathbf{\Theta}^{k}\right)$
is the constraint under the mean-field assumption \cite{Parisi_MeanField}.
The KL divergence in (\ref{eq:Problem_VBI}) is convex w.r.t. a single
variational distribution $q\left(\mathbf{\Theta}^{k}\right)$ after
fixing other variational distributions $q\left(\mathbf{\Theta}^{l}\right),\forall l\neq k$
\cite{ChengLei_Array_manifold}. Therefore, we can find a stationary
solution of (\ref{eq:Problem_VBI}) via optimizing each variational
distribution alternatively. And the optimal $q\left(\mathbf{\Theta}^{k}\right)$
is given by \cite{Tzikas_VBI}
\begin{equation}
q\bigl(\mathbf{\Theta}^{k}\bigr)=\frac{\exp\left(\left\langle \ln p\left(\mathbf{\Theta},\boldsymbol{y}\right)\right\rangle _{\Pi_{l\neq k}q\left(\mathbf{\Theta}^{l}\right)}\right)}{\int\exp\left(\left\langle \ln p\left(\mathbf{\Theta},\boldsymbol{y}\right)\right\rangle _{\Pi_{l\neq k}q\left(\mathbf{\Theta}^{l}\right)}\right)\textrm{d}\mathbf{\Theta}^{k}},\label{eq:optimal_q}
\end{equation}
where $\left\langle \cdot\right\rangle _{\Pi_{l\neq k}q\left(\mathbf{\Theta}^{l}\right)}$
represents an expectation w.r.t. $q\left(\mathbf{\Theta}^{l}\right),\forall l\neq k$.
Submitting the joint distribution in (\ref{eq:joint distribution})
into (\ref{eq:optimal_q}), the update of $q\left(\boldsymbol{x}\right)$
is derived as a complex Gaussian distribution with its mean and covariance
given by
\begin{equation}
\begin{aligned}\boldsymbol{\mu} & =\mathbf{\Sigma}\left\langle \gamma\right\rangle \mathbf{F}^{H}\boldsymbol{y},\\
\mathbf{\Sigma} & =\left(\left\langle \gamma\right\rangle \mathbf{F}^{H}\mathbf{F}+\textrm{diag}\left(\left\langle \rho\right\rangle \right)\right)^{-1}.
\end{aligned}
\label{eq:q(x)_matrix_inverse}
\end{equation}
Note that the computation of $\mathbf{\Sigma}$ involves a $Q\times Q$
dimensional matrix inverse, whose complexity is $\mathcal{O}\left(Q^{3}\right)$.
The computational overhead caused by the matrix inverse is very high
due to the deployment of thousands of antennas in XL-MIMO systems
($M$ is large and $Q>M$).

Inspired by an inverse-free Bayesian approach in \cite{Duan_IFSBL},
we develop a low-complexity IF-VBI algorithm that avoids the matrix
inverse via minimizing a relaxed KL divergence. Note that the proposed
IF-VBI is a variant of the Bayesian approach in \cite{Duan_IFSBL}.
The difference is that the IF-VBI is used to deal with the hierarchical
sparse prior in (\ref{eq:p(x,rou,s)}) but not the Laplace prior in
\cite{Duan_IFSBL}. The main idea of the IF-VBI is to construct a
relaxed KL divergence and then minimize it based on the majorization-minimization
(MM) framework \cite{Sun_MM}.

Specifically, a lower bound of the likelihood function can be found
by resorting to Lemma 1 in \cite{Duan_IFSBL},
\begin{align}
p\left(\boldsymbol{y}\mid\boldsymbol{x},\gamma\right)= & \left(\frac{\gamma}{\pi}\right)^{M}\exp\left(-\gamma\left\Vert \boldsymbol{y}-\mathbf{F}\boldsymbol{x}\right\Vert ^{2}\right)\nonumber \\
\geq & \left(\frac{\gamma}{\pi}\right)^{M}\exp\left(-\gamma g\left(\boldsymbol{x},\boldsymbol{z}\right)\right)\triangleq\mathrm{G}\left(\boldsymbol{y},\boldsymbol{x},\boldsymbol{z},\gamma\right),\label{eq:lower_bound_likelihood}
\end{align}
with $g\left(\boldsymbol{x},\boldsymbol{z}\right)$ given by
\begin{align}
g\left(\boldsymbol{x},\boldsymbol{z}\right)\triangleq & \left\Vert \boldsymbol{y}-\mathbf{F}\boldsymbol{z}\right\Vert ^{2}+T\left\Vert \boldsymbol{x}-\boldsymbol{z}\right\Vert ^{2}\nonumber \\
 & +2\Re\left\{ \left(\boldsymbol{x}-\boldsymbol{z}\right)^{H}\mathbf{F}^{H}\left(\mathbf{F}\boldsymbol{z}-\boldsymbol{y}\right)\right\} ,
\end{align}
where $T$ needs to satisfy $T\mathbf{I}_{Q}\succeq\mathbf{F}^{H}\mathbf{F}$.
And a good choice for $T$ is $\lambda_{\textrm{max}}\left(\mathbf{F}^{H}\mathbf{F}\right)$,
where $\lambda_{\textrm{max}}\left(\cdot\right)$ denotes the largest
eigenvalue of the given matrix.

Submitting (\ref{eq:lower_bound_likelihood}) into (\ref{eq:Problem_VBI}),
we construct a relaxed KL divergence as
\begin{equation}
\widetilde{\textrm{KL}}\left(q\parallel p\right)\triangleq\int q\left(\mathbf{\Theta}\right)\ln\frac{q\left(\mathbf{\Theta}\right)p\left(\boldsymbol{y}\right)}{\mathrm{G}\left(\boldsymbol{y},\boldsymbol{x},\boldsymbol{z},\gamma\right)p\left(\mathbf{\Theta}\right)}\mathrm{d}\boldsymbol{v},
\end{equation}
which is a upper bound of $\mathrm{KL}\left(q\parallel p\right)$.
Based on this, we employ the MM framework to minimize the relaxed
KL divergence w.r.t. $q\left(\mathbf{\Theta}^{k}\right),\forall k$
and $\boldsymbol{z}$. Specifically, in the majorization step, we
update each variational distribution alternatively after fixing $\boldsymbol{z}$.
In the minimization step, we minimize $\widetilde{\textrm{KL}}\left(q\parallel p\right)$
w.r.t. $\boldsymbol{z}$ given $q\left(\mathbf{\Theta}^{k}\right),\forall k$.
The IF-VBI estimator iterates between the majorization step and minimization
step until convergence. 

\subsubsection{Majorization step (update of $q\left(\mathbf{\Theta}^{k}\right),\forall k$)}

Using (\ref{eq:optimal_q}), $q\left(\boldsymbol{x}\right)$ is update
as
\begin{align}
\ln q\left(\boldsymbol{x}\right)\propto & \left\langle \ln\mathrm{G}\left(\boldsymbol{y},\boldsymbol{x},\boldsymbol{z},\gamma\right)\right\rangle _{q\left(\gamma\right)}+\left\langle \ln p\left(\boldsymbol{x}\mid\boldsymbol{\rho}\right)\right\rangle _{q\left(\boldsymbol{\rho}\right)}\nonumber \\
\propto & -\boldsymbol{x}^{H}\left(\left\langle \gamma\right\rangle T\mathbf{I}_{Q}+\textrm{diag}\left(\left\langle \boldsymbol{\rho}\right\rangle \right)\right)\boldsymbol{x}\nonumber \\
 & +2\Re\left\{ \boldsymbol{x}^{H}\left\langle \gamma\right\rangle \left(\mathbf{F}^{H}\left(\boldsymbol{y}-\mathbf{F}\boldsymbol{z}\right)+T\boldsymbol{z}\right)\right\} .
\end{align}
This is a complex Gaussian distribution with mean $\boldsymbol{\mu}$
and covariance $\mathbf{\Sigma}$ given by
\begin{equation}
\begin{aligned}\boldsymbol{\mu} & =\mathbf{\Sigma}\left\langle \gamma\right\rangle \left(\mathbf{F}^{H}\left(\boldsymbol{y}-\mathbf{F}\boldsymbol{z}\right)+T\boldsymbol{z}\right),\\
\mathbf{\Sigma} & =\left(\left\langle \gamma\right\rangle T\mathbf{I}_{Q}+\textrm{diag}\left(\left\langle \boldsymbol{\rho}\right\rangle \right)\right)^{-1}.
\end{aligned}
\label{eq:q(x)}
\end{equation}
Now $\mathbf{\Sigma}$ is calculated by a diagonal matrix inverse,
which is linear complexity.

The update of $q\left(\gamma\right)$ can be derived as
\begin{align}
\ln q\left(\gamma\right) & \propto\left\langle \ln\mathrm{G}\left(\boldsymbol{y},\boldsymbol{x},\boldsymbol{z},\gamma\right)\right\rangle _{q\left(\boldsymbol{x}\right)}+\ln p\left(\gamma\right)\nonumber \\
 & \propto\left(c-1+M\right)\ln\gamma-\left(d+\left\langle g\left(x,z\right)\right\rangle \right)\gamma,
\end{align}
with $\left\langle g\left(x,z\right)\right\rangle $ given by
\begin{align}
\left\langle g\left(x,z\right)\right\rangle = & \left\Vert \boldsymbol{y}-\mathbf{F}\boldsymbol{z}\right\Vert ^{2}+T\left\Vert \boldsymbol{\mu}-\boldsymbol{z}\right\Vert ^{2}+\textrm{Tr}\left(\mathbf{\Sigma}\right)\nonumber \\
 & +2\Re\left\{ \left(\boldsymbol{\mu}-\boldsymbol{z}\right)^{H}\mathbf{F}^{H}\left(\mathbf{F}\boldsymbol{z}-\boldsymbol{y}\right)\right\} ,
\end{align}
where $\textrm{Tr}\left(\cdot\right)$ is the trace of the given matrix.
Thus, we have 
\begin{equation}
q\left(\gamma\right)=\textrm{Ga}\left(\gamma;\tilde{c},\tilde{d}\right),
\end{equation}
where the parameters $\tilde{c}$ and $\tilde{d}$ are given by
\begin{equation}
\begin{aligned}\tilde{c} & =c+M,\\
\tilde{d} & =d+\left\langle g\left(x,z\right)\right\rangle .
\end{aligned}
\label{eq:q(gamma)}
\end{equation}
The variational distribution of $\boldsymbol{\rho}$ and $\boldsymbol{s}$
can be derived in a similar way. Please refer to subsection IV-D in
\cite{LiuAn_CE_Turbo_VBI} for the expression of $q\left(\boldsymbol{\rho}\right)$
and $q\left(\boldsymbol{s}\right)$.

\subsubsection{Minimization step (update of $\boldsymbol{z}$)}

Submitting $q\left(\mathbf{\Theta}\right)$ into $\widetilde{\textrm{KL}}\left(q\bigparallel p\right)$,
$\boldsymbol{z}$ can be updated as
\begin{equation}
\boldsymbol{z}^{\textrm{new}}=\arg\min_{\boldsymbol{z}}\left\langle -\ln\mathrm{G}\left(\boldsymbol{y},\boldsymbol{x},\boldsymbol{z},\gamma\right)\right\rangle _{q\left(\mathbf{\Theta}\right)}.\label{eq:probelm_minizization}
\end{equation}
Calculate the gradient of the function in (\ref{eq:probelm_minizization}),
i.e.,
\begin{equation}
\nabla_{\boldsymbol{z}}\left\langle -\ln\mathrm{G}\left(\boldsymbol{y},\boldsymbol{x},\boldsymbol{z},\gamma\right)\right\rangle _{q\left(\mathbf{\Theta}\right)}=\left\langle \gamma\right\rangle \left(\mathbf{F}^{H}\mathbf{F}-T\mathbf{I}_{Q}\right)\left(\boldsymbol{\mu}-\boldsymbol{z}\right).
\end{equation}
The function is minimized when the gradient becomes zero, i.e.,
\begin{equation}
\boldsymbol{z}^{\textrm{new}}=\boldsymbol{\mu}.\label{eq:z_new}
\end{equation}

Finally, the MAP estimate of $\boldsymbol{x}$ and the MMSE estimate
of $\boldsymbol{\gamma}$ are given by
\begin{equation}
\begin{aligned}\hat{\boldsymbol{x}} & =\boldsymbol{\mu},\\
\hat{\gamma} & =\frac{\tilde{c}}{\tilde{d}}.
\end{aligned}
\label{eq:MAP_x_gamma}
\end{equation}

\subsection{Structured EP (VR Detection Module)}

Given $\hat{\boldsymbol{x}}$, $\hat{\gamma}$, and $\hat{\boldsymbol{\Xi}}$,
the observation model in (\ref{eq:y_MAP}) is rewritten as $\boldsymbol{y}=\left[\mathbf{A}\left(\hat{\boldsymbol{\Xi}}\right)\varodot\mathbf{U}\left(\mathbf{V}\right)\right]\hat{\boldsymbol{x}}+\boldsymbol{w}.$
We employ the sub-array grouping method since a sub-array is the basic
unit keeping spatial stationary. Define the index set of the antennas
in sub-array $\left(k_{x},k_{z}\right)$ as $\mathbf{\Psi}_{k_{x},k_{z}}\subseteq\left\{ 1,\ldots,M\right\} $,
with its cardinal number given by $\left|\mathbf{\Psi}_{k_{x},k_{z}}\right|=N$.
Then, the received signal of sub-array $\left(k_{x},k_{z}\right)$
is expressed as
\begin{align}
\boldsymbol{y}_{k_{x},k_{z}}= & \left[\mathbf{A}_{k_{x},k_{z}}\left(\hat{\boldsymbol{\Xi}}\right)\varodot\left(\mathbf{1}_{N\times1}\hat{\boldsymbol{x}}^{T}\right)\right]\boldsymbol{v}_{k_{x},k_{z}}+\boldsymbol{w}_{k_{x},k_{z}},\label{eq:y_sub_array}
\end{align}
with $\boldsymbol{y}_{k_{x},k_{z}}\triangleq\left[y_{m}\right]_{m\in\mathbf{\Psi}_{k_{x},k_{z}}}$,
$\boldsymbol{w}_{k_{x},k_{z}}\triangleq\left[w_{m}\right]_{m\in\mathbf{\Psi}_{k_{x},k_{z}}}$,
and $\mathbf{A}_{k_{x},k_{z}}\left(\hat{\boldsymbol{\Xi}}\right)\triangleq\left[\mathbf{A}_{m}\left(\hat{\boldsymbol{\Xi}}\right)\right]_{m\in\mathbf{\Psi}_{k_{x},k_{z}}}$.
$\mathbf{A}_{m}\left(\hat{\boldsymbol{\Xi}}\right)$ denotes the $m\textrm{-th}$
row of $\mathbf{A}\left(\hat{\boldsymbol{\Xi}}\right)$, and $\boldsymbol{v}_{k_{x},k_{z}}\triangleq\left[v_{1,k_{x},k_{z}},\ldots,v_{Q,k_{x},k_{z}}\right]^{T}$
is the VR of sub-array $\left(k_{x},k_{z}\right)$. Moreover, we define
$\mathbf{H}_{k_{x},k_{z}}\triangleq\mathbf{A}_{k_{x},k_{z}}\left(\hat{\boldsymbol{\Xi}}\right)\odot\left(\mathbf{1}_{N\times1}\hat{\boldsymbol{x}}^{T}\right)$
to simplify the notation. 

Note that $\hat{\boldsymbol{x}}$ has only a few non-zero element,
which means that many columns of $\mathbf{H}_{k_{x},k_{z}}$ are all-zero
vectors. In this case, it is difficult to estimate $\boldsymbol{v}_{k_{x},k_{z}}$
from the observation $\boldsymbol{y}_{k_{x},k_{z}}$ under an ill-conditioned
sensing matrix $\mathbf{H}_{k_{x},k_{z}}$. We use a polar-domain
filtering method to address this problem. Specifically, we set a small
threshold $\varepsilon>0$ and compare each element of $\hat{\boldsymbol{x}}$
with the threshold.\footnote{The threshold $\varepsilon$ is chosen according to the noise power
in practice. A good choice for $\varepsilon$ is 2 to 3 times the
noise power.} Let $\mathbf{\Omega}\triangleq\left\{ q\mid\forall\left\Vert \hat{x}_{q}\right\Vert ^{2}>\varepsilon\right\} $
represent the index set of the elements with the energy larger than
the threshold. Then, we only retain the columns indexed by $\mathbf{\Omega}$
in $\mathbf{H}_{k_{x},k_{z}}$ and delete other columns that are close
to zero. In this case, the obtained sensing matrix, denoted by $\mathbf{\ddot{H}}_{k_{x},k_{z}}\in\mathbb{C}^{N\times\left|\mathbf{\Omega}\right|}$,
is well-conditioned, and the received signal model in (\ref{eq:y_sub_array})
is rewritten as
\begin{equation}
\boldsymbol{y}_{k_{x},k_{z}}=\mathbf{\ddot{H}}_{k_{x},k_{z}}\ddot{\boldsymbol{v}}_{k_{x},k_{z}}+\boldsymbol{w}_{k_{x},k_{z}},\label{eq:y_sub_array_new}
\end{equation}
where $\ddot{\boldsymbol{v}}_{k_{x},k_{z}}\triangleq\left[v_{q,k_{x},k_{z}}\right]_{q\in\mathbf{\Omega}}$.
Such a polar-domain filtering method also greatly reduces the complexity
of the algorithm due to $\left|\mathbf{\Omega}\right|\ll Q$. For
$q\notin\mathbf{\Omega}$, $\hat{x}_{q}$ is close to zero, which
indicates that there is no scatterer lying around the $q\textrm{-th}$
polar-domain grid point. Therefore, it is unnecessary to estimate
$v_{q,k_{x},k_{z}}$ for $q\notin\mathbf{\Omega}$, and we can simply
set them to be zero.

Since $\ddot{\boldsymbol{v}}_{k_{x},k_{z}},\forall k_{x},k_{z}$ is
real-valued, we reformulate the complex-valued model in (\ref{eq:y_sub_array_new})
into a real-valued one,
\begin{equation}
\tilde{\boldsymbol{y}}_{k_{x},k_{z}}=\mathbf{\tilde{H}}_{k_{x},k_{z}}\ddot{\boldsymbol{v}}_{k_{x},k_{z}}+\tilde{\boldsymbol{w}}_{k_{x},k_{z}},\label{eq:y_sub_array_EP}
\end{equation}
where $\tilde{\boldsymbol{y}}_{k_{x},k_{z}}\triangleq\left[\Re\left\{ \boldsymbol{y}_{k_{x},k_{z}}\right\} ^{T},\Im\left\{ \boldsymbol{y}_{k_{x},k_{z}}\right\} ^{T}\right]^{T}$,
and $\mathbf{\tilde{H}}_{k_{x},k_{z}}$ and $\tilde{\boldsymbol{w}}_{k_{x},k_{z}}$
are defined similarly. Based on the above linear observation model,
we develop a structured EP algorithm to compute the approximate posterior
of $\ddot{\boldsymbol{v}}_{k_{x},k_{z}},\forall k_{x},k_{z}$.

As shown in Fig. \ref{fig:The-alternating-MAP}, the structured EP
is a turbo framework that consists of two basic modules: Module A
is a LMMSE estimator that combines the observation information and
messages from Module B, while Module B is called the 2D-Markov-MMSE
estimator, which performs the MMSE estimation based on the 2D Markov
prior and messages from Module A. The two modules exchange messages
and work alternatively until convergence. 

\subsubsection{LMMSE in Module A}

In Module A, a Gaussian distribution is assumed as the prior for $\ddot{\boldsymbol{v}}_{k_{x},k_{z}}$,
denoted by $\mathcal{N}\left(\ddot{\boldsymbol{v}}_{k_{x},k_{z}};\boldsymbol{\alpha}_{k_{x},k_{z}}^{\textrm{A,pri}},\textrm{diag}\left(\boldsymbol{\beta}_{k_{x},k_{z}}^{\textrm{A,pri}}\right)\right)$,
where $\boldsymbol{\alpha}_{k_{x},k_{z}}^{\textrm{A,pri}}$ and $\boldsymbol{\beta}_{k_{x},k_{z}}^{\textrm{A,pri}}$
are extrinsic messages from module B. Based on the LMMSE estimator,
the posterior distribution is also a Gaussian distribution with its
mean and covariance given by
\begin{equation}
\begin{aligned}\mathbf{\Gamma}_{k_{x},k_{z}}^{\textrm{A,post}} & =\left(\hat{\gamma}\mathbf{\tilde{H}}_{k_{x},k_{z}}^{T}\mathbf{\tilde{H}}_{k_{x},k_{z}}+\textrm{diag}\left(1/\boldsymbol{\beta}_{k_{x},k_{z}}^{\textrm{A,pri}}\right)\right)^{-1},\\
\boldsymbol{\alpha}_{k_{x},k_{z}}^{\textrm{A,post}} & =\mathbf{\Gamma}_{k_{x},k_{z}}^{\textrm{A,post}}\left(\boldsymbol{\alpha}_{k_{x},k_{z}}^{\textrm{A,pri}}/\boldsymbol{\beta}_{k_{x},k_{z}}^{\textrm{A,pri}}+\hat{\gamma}\mathbf{\tilde{H}}_{k_{x},k_{z}}^{T}\tilde{\boldsymbol{y}}_{k_{x},k_{z}}\right).
\end{aligned}
\label{eq:A_post}
\end{equation}
Although the calculation of $\mathbf{\Gamma}_{k_{x},k_{z}}^{\textrm{A,post}}$
involves an $\left|\mathbf{\Omega}\right|\times\left|\mathbf{\Omega}\right|$
dimensional matrix inverse, its complexity is relatively low since
$\left|\mathbf{\Omega}\right|$ is small ($\left|\mathbf{\Omega}\right|$
is comparable to the number of channel paths). Then, the extrinsic
message passed from Module A to Module B is computed by \cite{Cespedes_EP}
\begin{equation}
\begin{aligned}\boldsymbol{\alpha}_{k_{x},k_{z}}^{\textrm{B,pri}} & =\boldsymbol{\beta}_{k_{x},k_{z}}^{\textrm{B,pri}}\left(\boldsymbol{\alpha}_{k_{x},k_{z}}^{\textrm{A,post}}/\boldsymbol{\beta}_{k_{x},k_{z}}^{\textrm{A,post}}-\boldsymbol{\alpha}_{k_{x},k_{z}}^{\textrm{A,pri}}/\boldsymbol{\beta}_{k_{x},k_{z}}^{\textrm{A,pri}}\right),\\
\boldsymbol{\beta}_{k_{x},k_{z}}^{\textrm{B,pri}} & =1/\left(1/\boldsymbol{\beta}_{k_{x},k_{z}}^{\textrm{A,post}}-1/\boldsymbol{\beta}_{k_{x},k_{z}}^{\textrm{A,pri}}\right),
\end{aligned}
\label{eq:A_ext}
\end{equation}
where $\boldsymbol{\beta}_{k_{x},k_{z}}^{\textrm{A,post}}=\textrm{diag}\left(\mathbf{\Gamma}_{k_{x},k_{z}}^{\textrm{A,post}}\right)$.

\subsubsection{Message passing in Module B}

A basic assumption is to model $\boldsymbol{\alpha}_{k_{x},k_{z}}^{\textrm{B,pri}}$
as an AWGN observation \cite{Cespedes_EP}:
\begin{equation}
\boldsymbol{\alpha}_{k_{x},k_{z}}^{\textrm{B,pri}}=\ddot{\boldsymbol{v}}_{k_{x},k_{z}}+\boldsymbol{z}_{k_{x},k_{z}},\label{eq:virtual AWGN}
\end{equation}
where $\boldsymbol{z}_{k_{x},k_{z}}\sim\mathcal{N}\left(0,\textrm{diag}\left(\boldsymbol{\beta}_{k_{x},k_{z}}^{\textrm{B,pri}}\right)\right)$
is the virtual equivalent noise. Such an assumption has been widely
used in EP \cite{Cespedes_EP} and message-passing-based algorithms
\cite{LiuAn_CE_Turbo_CS,Yuan_BGprior1}.

\begin{figure}[t]
\begin{centering}
\includegraphics[width=75mm]{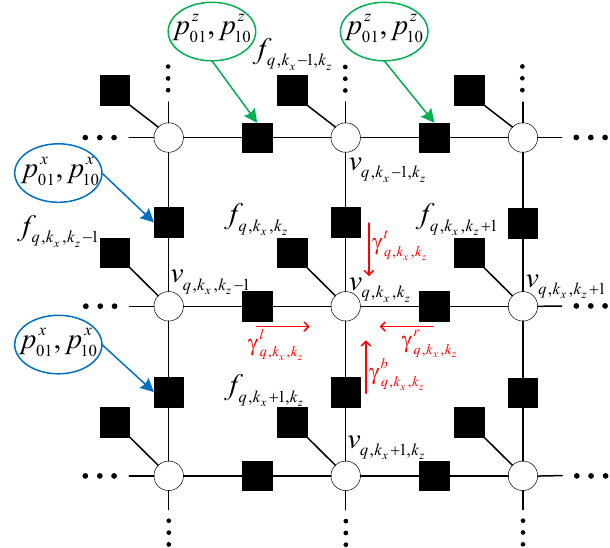}
\par\end{centering}
\caption{\label{fig:Factor-graph-Module-B}The internal structure of each sub-graph
$\mathcal{G}_{q},q\in\mathbf{\Omega}$.}
\end{figure}

Denote the collection of measurements and variables in (\ref{eq:virtual AWGN})
as $\boldsymbol{\alpha}^{\textrm{B,pri}}\triangleq\left\{ \boldsymbol{\alpha}_{k_{x},k_{z}}^{\textrm{B,pri}}\mid\forall k_{x},k_{z}\right\} $
and $\ddot{\boldsymbol{v}}\triangleq\left\{ \ddot{\boldsymbol{v}}_{k_{x},k_{z}}\mid\forall k_{x},k_{z}\right\} $,
respectively, then the joint distribution of $\boldsymbol{\alpha}^{\textrm{B,pri}}$
and $\ddot{\boldsymbol{v}}$ is expressed as
\begin{equation}
p\left(\boldsymbol{\alpha}^{\textrm{B,pri}},\ddot{\boldsymbol{v}}\right)=\prod_{q\in\mathbf{\Omega}}\biggl(p\left(\mathbf{V}_{q}\right)\prod_{k_{x}=1}^{K_{x}}\prod_{k_{z}=1}^{K_{z}}p\left(\alpha_{q,k_{x},k_{z}}^{\textrm{B,pri}}\mid v_{q,k_{x},k_{z}}\right)\biggr),\label{eq:joint_PDF_MuduleB}
\end{equation}
where $p\left(\mathbf{V}_{q}\right)$ is the 2D Markov prior given
in (\ref{eq:2D-MM}), and $v_{q,k_{x},k_{z}}$ and $\alpha_{q,k_{x},k_{z}}^{\textrm{B,pri}}$
denote the element of $\ddot{\boldsymbol{v}}_{k_{x},k_{z}}$ and $\boldsymbol{\alpha}_{k_{x},k_{z}}^{\textrm{B,pri}}$,
respectively, for $q\in\mathbf{\Omega}$. Note that in Module B, $\mathbf{V}_{q},q\in\mathbf{\Omega}$
are treated as binary variables with the 2D Markov prior. Even though
$\ddot{\boldsymbol{v}}_{k_{x},k_{z}},\forall k_{x},k_{z}$ are assumed
to be Gaussian distribution in Module A, they will be ``projected''
back to binary variables in Module B each iteration. And thus, the
Gaussian approximation error in Module A can be well controlled, as
verified by simulations. According to (\ref{eq:joint_PDF_MuduleB}),
the factor graph of the joint distribution consists of $\left|\mathbf{\Omega}\right|$
independent sub-graphs, and each sub-graph, denoted by $\mathcal{G}_{q},q\in\mathbf{\Omega}$,
has the same internal structure. The sub-graph $\mathcal{G}_{q}$
is presented in Fig. \ref{fig:Factor-graph-Module-B}, where the factor
nodes are defined as 
\[
f_{q,k_{x},k_{z}}\triangleq\mathcal{N}\left(v_{q,k_{x},k_{z}};\alpha_{q,k_{x},k_{z}}^{\textrm{B,pri}},\beta_{q,k_{x},k_{z}}^{\textrm{B,pri}}\right),q\in\mathbf{\Omega},\forall k_{x},k_{z}.
\]
We obey the sum-product rule \cite{Ksch_sum-product} to perform message
passing over each sub-graph. Consider a variable node $v_{q,k_{x},k_{z}}$,
whose input messages from left, right, top, and bottom factor nodes
are denoted by $\mathcal{V}_{q,k_{x},k_{z}}^{l}$, $\mathcal{V}_{q,k_{x},k_{z}}^{r}$,
$\mathcal{V}_{q,k_{x},k_{z}}^{t}$, and $\mathcal{V}_{q,k_{x},k_{z}}^{b}$,
respectively. The input messages are calculated as
\begin{align}
\mathcal{V}_{q,k_{x},k_{z}}^{d}= & \gamma_{q,k_{x},k_{z}}^{d}\delta\left(v_{q,k_{x},k_{z}}-1\right)\nonumber \\
+ & \left(1-\gamma_{q,k_{x},k_{z}}^{d}\right)\delta\left(v_{q,k_{x},k_{z}}\right),d\in\left\{ l,r,t,b\right\} ,\label{eq:message_lrtb}
\end{align}
with $\gamma_{q,k_{x},k_{z}}^{d},d\in\left\{ l,r,t,b\right\} $ given
at the top of the next page, where $\pi_{q,k_{x},k_{z}}^{in,1}=\mathcal{N}\left(1;\alpha_{q,k_{x},k_{z}}^{\textrm{B,pri}},\beta_{q,k_{x},k_{z}}^{\textrm{B,pri}}\right)$
and $\pi_{q,k_{x},k_{z}}^{in,0}=\mathcal{N}\left(0;\alpha_{q,k_{x},k_{z}}^{\textrm{B,pri}},\beta_{q,k_{x},k_{z}}^{\textrm{B,pri}}\right)$.

Then, the output message passed from the variable node $v_{q,k_{x},k_{z}}$
to the factor node $f_{q,k_{x},k_{z}}$ is expressed as
\begin{figure*}[tbh]
\begin{subequations}
\begin{align}
\gamma_{q,k_{x},k_{z}}^{l}= & \frac{p_{11}^{z}\pi_{q,k_{x},k_{z}-1}^{in,1}\prod_{d\in\left\{ l,t,b\right\} }\gamma_{q,k_{x},k_{z}-1}^{d}+p_{10}^{z}\pi_{q,k_{x},k_{z}-1}^{in,0}\prod_{d\in\left\{ l,t,b\right\} }\left(1-\gamma_{q,k_{x},k_{z}-1}^{d}\right)}{\pi_{q,k_{x},k_{z}-1}^{in,1}\prod_{d\in\left\{ l,t,b\right\} }\gamma_{q,k_{x},k_{z}-1}^{d}+\pi_{q,k_{x},k_{z}-1}^{in,0}\prod_{d\in\left\{ l,t,b\right\} }\left(1-\gamma_{q,k_{x},k_{z}-1}^{d}\right)},\\
\gamma_{q,k_{x},k_{z}}^{r}= & \frac{p_{11}^{z}\pi_{q,k_{x},k_{z}+1}^{in,1}\prod_{d\in\left\{ r,t,b\right\} }\gamma_{q_{i},k_{x},k_{z}+1}^{d}+p_{10}^{z}\pi_{q,k_{x},k_{z}+1}^{in,0}\prod_{n\in\left\{ r,t,b\right\} }\left(1-\gamma_{q,k_{x},k_{z}+1}^{d}\right)}{\left(p_{11}^{z}+p_{01}^{z}\right)\pi_{q,k_{x},k_{z}+1}^{in,1}\prod_{d\in\left\{ r,t,b\right\} }\gamma_{q,k_{x},k_{z}+1}^{d}+\left(p_{00}^{z}+p_{10}^{z}\right)\pi_{q,k_{x},k_{z}+1}^{in,0}\prod_{d\in\left\{ r,t,b\right\} }\left(1-\gamma_{q,k_{x},k_{z}+1}^{d}\right)},\\
\gamma_{q,k_{x},k_{z}}^{t}= & \frac{p_{11}^{x}\pi_{q,k_{x}-1,k_{z}}^{in,1}\prod_{d\in\left\{ l,r,t\right\} }\gamma_{q,k_{x}-1,k_{z}}^{d}+p_{10}^{x}\pi_{q,k_{x}-1,k_{z}}^{in,0}\prod_{d\in\left\{ l,r,t\right\} }\left(1-\gamma_{q,k_{x}-1,k_{z}}^{d}\right)}{\pi_{q,k_{x}-1,k_{z}}^{in,1}\prod_{d\in\left\{ l,r,t\right\} }\gamma_{q,k_{x}-1,k_{z}}^{d}+\pi_{q,k_{x}-1,k_{z}}^{in,0}\prod_{d\in\left\{ l,r,t\right\} }\left(1-\gamma_{q,k_{x}-1,k_{z}}^{d}\right)},\\
\gamma_{q,k_{x},k_{z}}^{b}= & \frac{p_{11}^{x}\pi_{q,k_{x}+1,k_{z}}^{in,1}\prod_{d\in\left\{ l,r,b\right\} }\gamma_{q,k_{x}+1,k_{z}}^{d}+p_{10}^{x}\pi_{q,k_{x}+1,k_{z}}^{in,0}\prod_{d\in\left\{ l,r,b\right\} }\left(1-\gamma_{q,k_{x}+1,k_{z}}^{d}\right)}{\left(p_{11}^{x}+p_{01}^{x}\right)\pi_{q,k_{x}+1,k_{z}}^{in,1}\prod_{d\in\left\{ l,r,b\right\} }\gamma_{q,k_{x}+1,k_{z}}^{d}+\left(p_{00}^{x}+p_{10}^{x}\right)\pi_{q,k_{x}+1,k_{z}}^{in,0}\prod_{d\in\left\{ l,r,b\right\} }\left(1-\gamma_{q,k_{x}+1,k_{z}}^{d}\right)},
\end{align}
\end{subequations}

\rule[0.5ex]{1\textwidth}{1pt}
\end{figure*}
\begin{align}
\mathcal{V}_{q,k_{x},k_{z}}^{out}= & \pi_{q,k_{x},k_{z}}^{out}\delta\left(v_{q,k_{x},k_{z}}-1\right)\nonumber \\
 & +\left(1-\pi_{q,k_{x},k_{z}}^{out}\right)\delta\left(v_{q,k_{x},k_{z}}\right),\label{eq:message_out}
\end{align}
with
\[
\pi_{q,k_{x},k_{z}}^{out}=\tfrac{\prod_{d\in\left\{ l,r,t,b\right\} }\gamma_{q,k_{x},k_{z}}^{d}}{\prod_{d\in\left\{ l,r,t,b\right\} }\gamma_{q,k_{x},k_{z}}^{d}+\prod_{d\in\left\{ l,r,t,b\right\} }\left(1-\gamma_{q,k_{x},k_{z}}^{d}\right)}.
\]
Now, we can obtain the posterior probability of $v_{q,k_{x},k_{z}}$
as
\[
\hat{p}\left(v_{q,k_{x},k_{z}}=1\right)=\tfrac{\pi_{q,k_{x},k_{z}}^{in,1}\pi_{q,k_{x},k_{z}}^{out}}{\pi_{q,k_{x},k_{z}}^{in,1}\pi_{q,k_{x},k_{z}}^{out}+\pi_{q,k_{x},k_{z}}^{in,0}\left(1-\pi_{q,k_{x},k_{z}}^{out}\right)}.
\]
Based on this, the posterior mean and variance of $v_{q,k_{x},k_{z}},q\in\mathbf{\Omega},\forall k_{x},k_{z}$
are computed by
\begin{equation}
\begin{aligned}\alpha_{q,k_{x},k_{z}}^{\textrm{B,post}}= & \sum_{a\in\left\{ 0,1\right\} }\hat{p}\left(v_{q,k_{x},k_{z}}=a\right),\\
\beta_{q,k_{x},k_{z}}^{\textrm{B,post}}= & \sum_{a\in\left\{ 0,1\right\} }\left(a-\alpha_{q,k_{x},k_{z}}^{\textrm{B,post}}\right)^{2}\hat{p}\left(v_{q,k_{x},k_{z}}=a\right).
\end{aligned}
\label{eq:B_post}
\end{equation}
Define $\boldsymbol{\alpha}_{k_{x},k_{z}}^{\textrm{B,post}}\triangleq\left[\alpha_{q,k_{x},k_{z}}^{\textrm{B,post}}\right]_{q\in\mathbf{\Omega}}$
and $\boldsymbol{\beta}_{k_{x},k_{z}}^{\textrm{B,post}}\triangleq\left[\beta{}_{q,k_{x},k_{z}}^{\textrm{B,post}}\right]_{q\in\mathbf{\Omega}}$,
the extrinsic message passed from Module B to Module A is given by
\begin{equation}
\begin{aligned}\boldsymbol{\bar{\alpha}}_{k_{x},k_{z}}^{\textrm{A,pri}} & =\boldsymbol{\bar{\beta}}_{k_{x},k_{z}}^{\textrm{A,pri}}\left(\boldsymbol{\alpha}_{k_{x},k_{z}}^{\textrm{B,post}}/\boldsymbol{\beta}_{k_{x},k_{z}}^{\textrm{B,post}}-\boldsymbol{\alpha}_{k_{x},k_{z}}^{\textrm{B,pri}}/\boldsymbol{\beta}_{k_{x},k_{z}}^{\textrm{B,pri}}\right),\\
\bar{\boldsymbol{\beta}}_{k_{x},k_{z}}^{\textrm{A,pri}} & =1/\left(1/\boldsymbol{\beta}_{k_{x},k_{z}}^{\textrm{B,post}}-1/\boldsymbol{\beta}_{k_{x},k_{z}}^{\textrm{B,pri}}\right).
\end{aligned}
\label{eq:B_ext}
\end{equation}
Notably, the update of the extrinsic variance will result in a negative
$\bar{\beta}_{q,k_{x},k_{z}}^{\textrm{A,pri}}$ when $\beta_{q,k_{x},k_{z}}^{\textrm{B,post}}>\beta_{q,k_{x},k_{z}}^{\textrm{B,pri}}$,
which is unreasonable. If the calculation of the posterior is exact,
$\beta_{q,k_{x},k_{z}}^{\textrm{B,post}}>\beta_{q,k_{x},k_{z}}^{\textrm{B,pri}}$
will never happen since the posterior variance should be less than
the prior variance. However, since sum-product is not exact for the
2D Markov prior whose factor graph has loops, this case may occur
occasionally. In this case, we simply retain the previous values for
$\alpha_{q,k_{x},k_{z}}^{\textrm{A,pri}}$and $\beta_{q,k_{x},k_{z}}^{\textrm{A,pri}}$.
Such an approach is also employed in the conventional EP algorithm
to ensure numerical stability \cite{Cespedes_EP}. In addition, a
damping technique is usually used to smooth the update of extrinsic
message \cite{Cespedes_EP,Pratik_EP2}:
\begin{equation}
\begin{aligned}\boldsymbol{\alpha}_{k_{x},k_{z}}^{\textrm{A,pri}}= & \eta\bar{\boldsymbol{\alpha}}_{k_{x},k_{z}}^{\textrm{A,pri}}+\left(1-\eta\right)\boldsymbol{\alpha}_{k_{x},k_{z}}^{\textrm{A,pri}},\\
\boldsymbol{\beta}_{k_{x},k_{z}}^{\textrm{A,pri}}= & \eta\bar{\boldsymbol{\beta}}_{k_{x},k_{z}}^{\textrm{A,pri}}+\left(1-\eta\right)\boldsymbol{\beta}_{k_{x},k_{z}}^{\textrm{A,pri}},
\end{aligned}
\label{eq:damping}
\end{equation}
where $\eta\in\left(0,1\right)$ is a damping factor. The damping
in (\ref{eq:damping}) makes the structured EP more robust with improved
stability and convergence properties.

Finally, we update the MAP estimate of $\hat{\mathbf{V}}_{q},q\in\mathbf{\Omega}$
by a hard decision as
\begin{equation}
\hat{v}_{q,k_{x},k_{z}}=\begin{cases}
1, & \hat{p}\left(v_{q,k_{x},k_{z}}=1\right)\geq0.5,\\
0, & \hat{p}\left(v_{q,k_{x},k_{z}}=1\right)<0.5.
\end{cases},\forall k_{x},k_{z}.\label{eq:MAP_V}
\end{equation}

\subsection{Gradient Ascent (Grid Update Module)}

Based on $\hat{\boldsymbol{x}}$ and $\hat{\gamma}$ obtained in (\ref{eq:MAP_x_gamma})
and $\hat{\mathbf{V}}$ obtained in (\ref{eq:MAP_V}), the logarithmic
posterior function is expressed as
\begin{align}
\mathcal{L}\left(\boldsymbol{\Xi}\right) & \triangleq\ln p\left(\boldsymbol{y}\mid\hat{\mathbf{V}},\hat{\boldsymbol{x}},\hat{\gamma};\boldsymbol{\Xi}\right)+\ln\left(\hat{\mathbf{V}},\hat{\boldsymbol{x}},\hat{\gamma}\right)-\ln\left(\boldsymbol{y}\right)\nonumber \\
 & =-\hat{\gamma}\left\Vert \boldsymbol{y}-\left[\mathbf{A}\left(\boldsymbol{\Xi}\right)\varodot\mathbf{U}\left(\hat{\mathbf{V}}\right)\right]\hat{\boldsymbol{x}}\right\Vert ^{2}+C,
\end{align}
where $C$ is a constant. It is difficult to find the optimal $\boldsymbol{\boldsymbol{\Xi}}$
that maximizes $\mathcal{L}\left(\boldsymbol{\Xi}\right)$ since $\mathcal{L}\left(\boldsymbol{\Xi}\right)$
is non-concave w.r.t. $\boldsymbol{\Xi}$. In this case, a gradient
ascent approach is usually employed to update $\boldsymbol{\Xi}$.
Specifically, in the $i\textrm{-th}$ iteration, the angle and distance
parameters are updated as
\begin{align}
\hat{\boldsymbol{\vartheta}}^{\left(i\right)} & =\hat{\boldsymbol{\vartheta}}^{\left(i-1\right)}+\epsilon_{1}^{\left(i\right)}\nabla_{\boldsymbol{\vartheta}}\mathcal{L}\left(\boldsymbol{\vartheta},\boldsymbol{\hat{\phi}}^{\left(i-1\right)},\hat{\boldsymbol{r}}^{\left(i-1\right)}\right)\mid_{\boldsymbol{\vartheta}=\hat{\boldsymbol{\vartheta}}^{\left(i-1\right)}},\nonumber \\
\boldsymbol{\hat{\phi}}^{\left(i\right)} & =\boldsymbol{\hat{\phi}}^{\left(i-1\right)}+\epsilon_{2}^{\left(i\right)}\nabla_{\boldsymbol{\phi}}\mathcal{L}\left(\hat{\boldsymbol{\vartheta}}^{\left(i\right)},\boldsymbol{\hat{\phi}},\hat{\boldsymbol{r}}^{\left(i-1\right)}\right)\mid_{\boldsymbol{\phi}=\boldsymbol{\hat{\phi}}^{\left(i-1\right)}},\nonumber \\
\frac{1}{\boldsymbol{\hat{r}}^{\left(i\right)}} & =\frac{1}{\boldsymbol{\hat{r}}^{\left(i-1\right)}}+\epsilon_{3}^{\left(i\right)}\nabla_{\boldsymbol{r}}\mathcal{L}\left(\hat{\boldsymbol{\vartheta}}^{\left(i\right)},\boldsymbol{\hat{\phi}}^{\left(i\right)},\boldsymbol{r}\right)\mid_{\boldsymbol{r}=\hat{\boldsymbol{r}}^{\left(i-1\right)}},\label{eq:gradient_update}
\end{align}
where $\epsilon_{1}^{\left(i\right)}$, $\epsilon_{2}^{\left(i\right)}$,
and $\epsilon_{3}^{\left(i\right)}$ are step sizes determined by
the Armijo rule. Note that we update $\frac{1}{\hat{\boldsymbol{r}}}$
instead of $\hat{\boldsymbol{r}}$ since $\frac{1}{r}$ is uniformly
sampled in the polar domain \cite{Cui_polar_grid}.

\subsection{Complexity Analysis}

We summarize the proposed alternating MAP framework in Algorithm \ref{Alternating_MAP}.
In the channel estimation module, the complicated matrix inverse is
avoid, and only some matrix-vector product and diagonal matrix inverse
are needed, whose complexity is $\mathcal{O}\left(MQ\right)$ per
iteration. In the VR detection module, the complexity of the LMMSE
estimator in Module A is dominated by $K$ small-scale matrix inverse
operations, whose complexity is $\mathcal{O}\left(K\left|\mathbf{\Omega}\right|^{3}\right)$
per iteration. And the message passing in Module B is linear complexity.
In the grid update module, the complexity of the gradient calculation
is $\mathcal{O}\left(Q^{2}\right)$. Denote the inner iteration number
of the IF-VBI and the structured EP as $I_{1}$ and $I_{2}$, respectively,
the overall complexity of the alternating MAP is $\mathcal{O}\left(I_{1}MQ+I_{2}K\left|\mathbf{\Omega}\right|^{3}+Q^{2}\right)$
per outer iteration.
\begin{algorithm}[tbh]
\begin{singlespace}
{\small{}\caption{\label{Alternating_MAP}The alternating MAP algorithm}
}{\small\par}

\textbf{Input:} $\boldsymbol{y}$, initial grid $\boldsymbol{\bar{\Xi}}$,
inner iteration number $I_{1},I_{2}$, outer iteration number $I_{out}$.

\textbf{Output:} $\hat{\boldsymbol{x}}$, $\hat{\mathbf{V}}$, and
$\hat{\boldsymbol{\Xi}}$.

\begin{algorithmic}[1]

\STATE Initialize: $\hat{\mathbf{V}}_{q}=\mathbf{1}_{K_{x}\times K_{z}},\forall q$,
$\hat{\boldsymbol{\Xi}}=\bar{\mathbf{\Xi}}$.

\FOR{${\color{blue}{\color{black}i_{out}=1,\cdots,I_{out}}}$}

\STATE\textbf{\% Channel estimation module: IF-VBI estimator}

\STATE Initialize: $T=\lambda_{\textrm{max}}\left(\mathbf{F}^{H}\mathbf{F}\right)$,
$\boldsymbol{z}=\mathbf{F}^{H}\boldsymbol{y}$.

\FOR{${\color{blue}{\color{black}i_{1}=1,\cdots,I_{1}}}$}

\STATE \textbf{Majorization:} update each variational distribution
$q\left(\mathbf{\Theta}^{k}\right)$, using (\ref{eq:q(x)}) and (\ref{eq:q(gamma)}).

\STATE \textbf{Minimization:} update the parameter $\boldsymbol{z}$,
using (\ref{eq:z_new}).

\ENDFOR

\STATE Update $\hat{\boldsymbol{x}}$ and $\hat{\gamma}$, using
(\ref{eq:MAP_x_gamma}).

\STATE\textbf{\% VR detection module: structured EP}

\FOR{${\color{blue}{\color{black}i_{2}=1,\cdots,I_{2}}}$}

\STATE\textbf{\% Module A: LMMSE estimator}

\STATE Update $\boldsymbol{\alpha}_{k_{x},k_{z}}^{\textrm{A,post}}$
and $\boldsymbol{\beta}_{k_{x},k_{z}}^{\textrm{A,post}}$, using (\ref{eq:A_post}).

\STATE Calculate the extrinsic message from Module A to B based on
(\ref{eq:A_ext}). 

\STATE\textbf{\% Module B: 2D-Markov-MMSE estimator}

\STATE Perform message passing over each sub-graph, using (\ref{eq:message_lrtb})
and (\ref{eq:message_out}).

\STATE Update $\boldsymbol{\alpha}_{k_{x},k_{z}}^{\textrm{B,post}}$
and $\boldsymbol{\beta}_{k_{x},k_{z}}^{\textrm{B,post}}$, using (\ref{eq:B_post}).

\STATE Calculate the extrinsic message from Module B to A based on
(\ref{eq:B_ext}) and (\ref{eq:damping}). 

\ENDFOR

\STATE Update $\hat{\mathbf{V}}$, using (\ref{eq:MAP_V}).

\STATE\textbf{\% Grid update module: gradient ascent}

\STATE Update $\hat{\boldsymbol{\Xi}}$ ($\hat{\boldsymbol{\vartheta}},\boldsymbol{\hat{\phi}},\hat{\boldsymbol{r}}$),
using (\ref{eq:gradient_update}).

\ENDFOR

\STATE Output $\hat{\boldsymbol{x}}$, $\hat{\mathbf{V}}$, and $\hat{\boldsymbol{\Xi}}$.

\end{algorithmic}
\end{singlespace}
\end{algorithm}

\section{Simulation Results}

In this section, we evaluate the performance of our proposed method
through adequate numerical simulations. Some baselines and the proposed
method are summarized below.
\begin{figure*}[t]
\centering{}%
\begin{minipage}[t]{0.45\textwidth}%
\begin{center}
\includegraphics[clip,width=65mm]{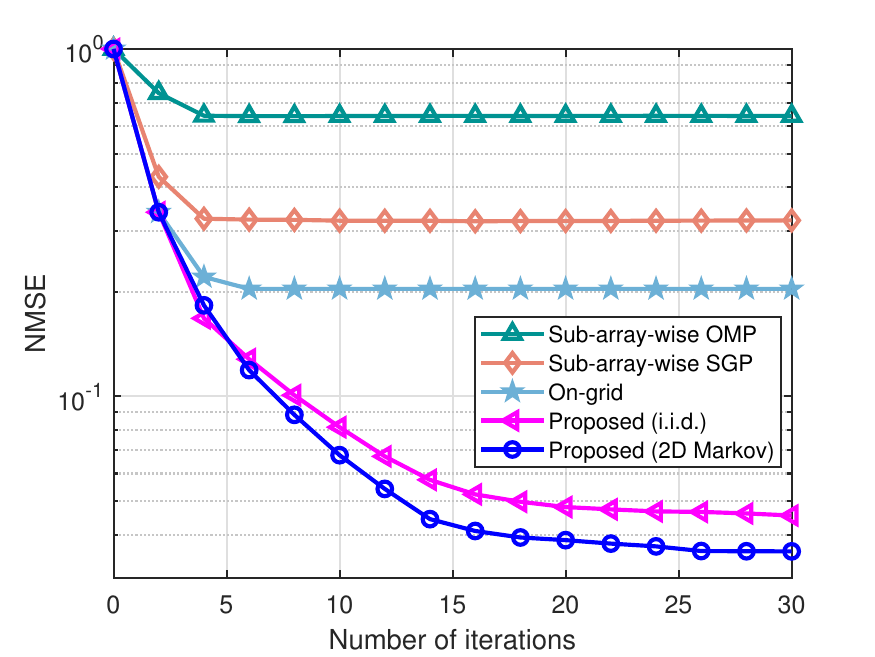}
\par\end{center}
\begin{center}
\vspace{-10mm}
\par\end{center}
\caption{\textcolor{blue}{\label{fig:Convergence_NMSE}}Convergence behavior:
NMSE of channel estimation with respect to the number of iterations
when $\textrm{SNR}=-4\ \textrm{dB}$.}
\end{minipage}\hfill{}%
\begin{minipage}[t]{0.45\textwidth}%
\begin{center}
\includegraphics[clip,width=65mm]{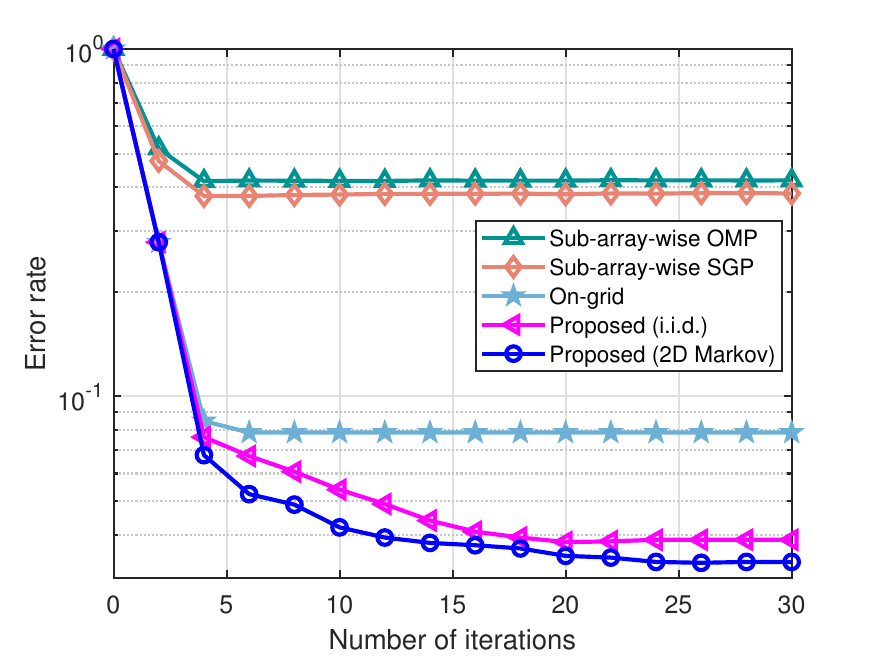}
\par\end{center}
\begin{center}
\vspace{-10mm}
\par\end{center}
\caption{\textcolor{blue}{\label{fig:Convergence_error_rate}}Convergence behavior:
error rate of VR detection with respect to the number of iterations
when $\textrm{SNR}=-4\ \textrm{dB}$.}
\end{minipage}
\end{figure*}

\begin{itemize}
\item \textbf{Sub-array-wise OMP \cite{HanYu_VR,Cui_polar_grid,Chen_NS_CE}:}
Each sub-array uses the OMP algorithm to estimate each sub-channel
independently and updates the polar-domain grid parameters via gradient
ascent. 
\item \textbf{Sub-array-wise stochastic gradient pursuit (SGP) \cite{Lin_SGP}:
}Each sub-array employs the SGP algorithm to estimate each sub-channel
independently and refines the polar-domain grid parameters via gradient
ascent.
\item \textbf{On-grid: }It is the proposed alternating MAP algorithm based
on the on-grid model, i.e., the polar-domain grid is fixed.
\item \textbf{Proposed (i.i.d.):} It is the proposed algorithm with the
i.i.d. Bernoulli prior for VRs. In this case, the VR detection module
is the conventional EP algorithm \cite{Cespedes_EP}.
\item \textbf{Proposed (2D Markov):} It is the proposed algorithm with the
2D Markov prior for VRs. 
\item \textbf{Genie-aided:} It is the proposed algorithm when the distance
and angle parameters are assumed to be known perfectly. And thus,
it is a performance upper bound for the proposed method.
\end{itemize}
The parameters of the considered XL-MIMO system are set as follows:
the number of the UPA antennas is $M_{x}=256$ and $M_{z}=8$; the
UPA is partitioned into $K_{x}\times K_{z}=16\times2$ small-scale
sub-UPAs, and the number of the sub-UPA antennas is $N_{x}=16$ and
$N_{z}=4$; the carrier frequency is $30\ \textrm{GHz}$, while the
Rayleigh distance is $325.37\ \textrm{m}$. We consider a multipath
channel model, where the number of paths is set to $L=4$. The whole
channel is spatial non-stationary, while each sub-channel corresponding
to each sub-array is assumed to be spatial stationary. The visibility
probability of the scatterers is $\kappa=0.5$, and the VR of the
scatterers concentrates on a few clusters. We use the ``COST 2100
Channel Model'' toolbox to generate the spatial non-stationary XL-MIMO
channel \cite{COST_2100_Model}. The normalized mean square error
(NMSE) is used as the performance metric for channel estimation. And
the error rate is used to measure the performance of VR detection,
which is defined as
\begin{equation}
\textrm{Error\ rate}\triangleq\frac{\sum_{l=1}^{L}\left|\mathbf{V}_{l}-\hat{\mathbf{V}}_{q_{l}}\right|}{K_{x}K_{z}L}\leq1,
\end{equation}
where $q_{l}$ is the index of the polar-domain grid point nearest
to the true position of scatterer $l$.

\subsection{Convergence Behavior}

In Fig. \ref{fig:Convergence_NMSE} and Fig. \ref{fig:Convergence_error_rate},
we compare the convergence behavior of different methods in terms
of channel estimation NMSE and VR detection error rate, respectively.
The convergence speed of the sub-array-wise OMP and SGP is very quick,
but both of them converge to a very poor stationary point. By contrast,
the on-grid method (the proposed algorithm based on the on-grid model)
can find a better stationary point than the sub-array-wise methods
and achieve similar convergence speed. Besides, the convergence behavior
of the proposed algorithm with the i.i.d. prior and the 2D Markov
prior is similar. Both of them converge within $20$ iterations, and
they achieve better steady-state performance than the on-grid method,
which shows the advantage of the designed dynamic polar-domain grid
over the fixed grid. Moreover, the proposed algorithm with the 2D
Markov prior has a performance gain over the same algorithm with the
i.i.d. prior after convergence. This is because the 2D Markov prior
can fully exploit the 2D clustered sparsity of VRs to enhance the
performance.
\begin{figure*}[t]
\centering{}%
\begin{minipage}[t]{0.45\textwidth}%
\begin{center}
\includegraphics[clip,width=65mm]{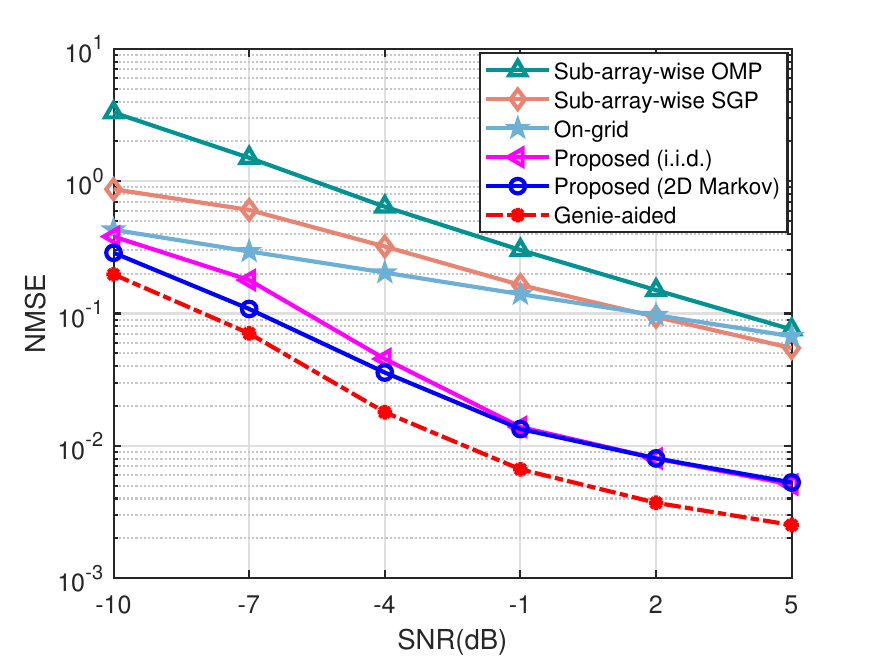}
\par\end{center}
\begin{center}
\vspace{-10mm}
\par\end{center}
\caption{\textcolor{blue}{\label{fig:NMSE_SNR}}NMSE of channel estimation
versus SNR.}
\end{minipage}\hfill{}%
\begin{minipage}[t]{0.45\textwidth}%
\begin{center}
\includegraphics[clip,width=65mm]{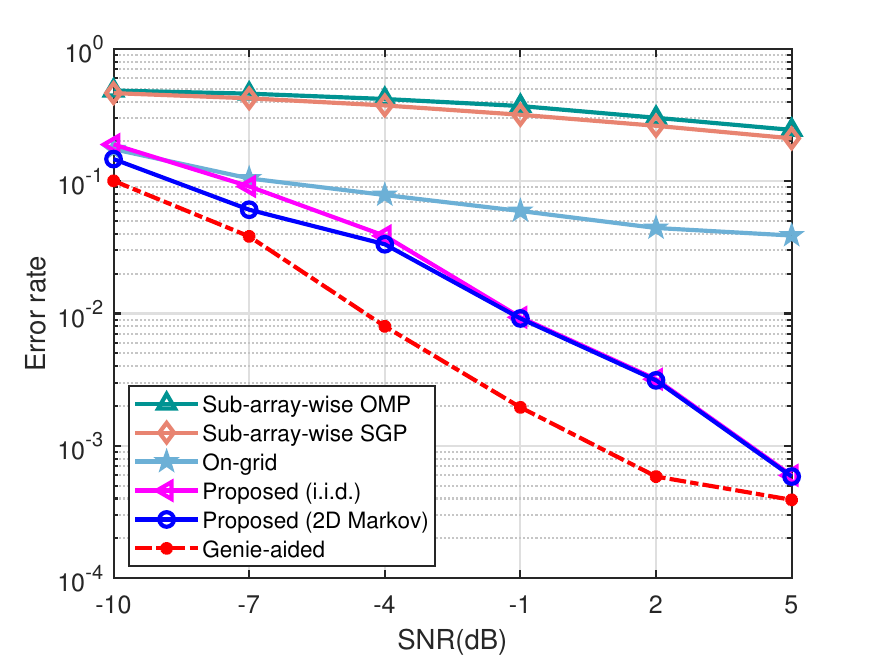}
\par\end{center}
\begin{center}
\vspace{-10mm}
\par\end{center}
\caption{\textcolor{blue}{\label{fig:Error_rate_SNR}}Error rate of VR detection
versus SNR.}
\end{minipage}
\end{figure*}
\begin{figure*}[t]
\centering{}%
\begin{minipage}[t]{0.45\textwidth}%
\begin{center}
\includegraphics[clip,width=65mm]{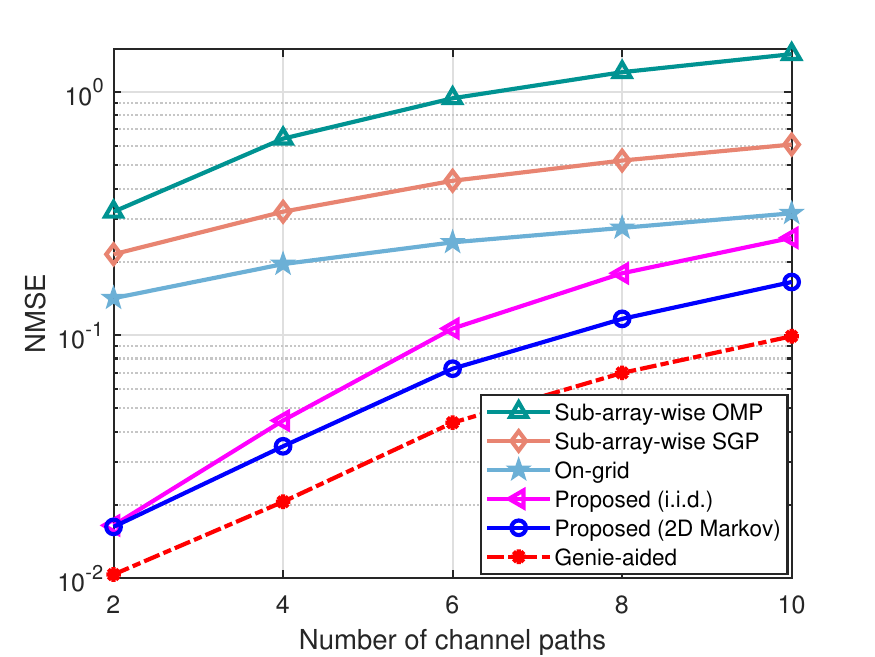}
\par\end{center}
\begin{center}
\vspace{-10mm}
\par\end{center}
\caption{\textcolor{blue}{\label{fig:NMSE_nPath}}NMSE of channel estimation
versus the number of channel paths when $\textrm{SNR}=-4\ \textrm{dB}$.}
\end{minipage}\hfill{}%
\begin{minipage}[t]{0.45\textwidth}%
\begin{center}
\includegraphics[clip,width=65mm]{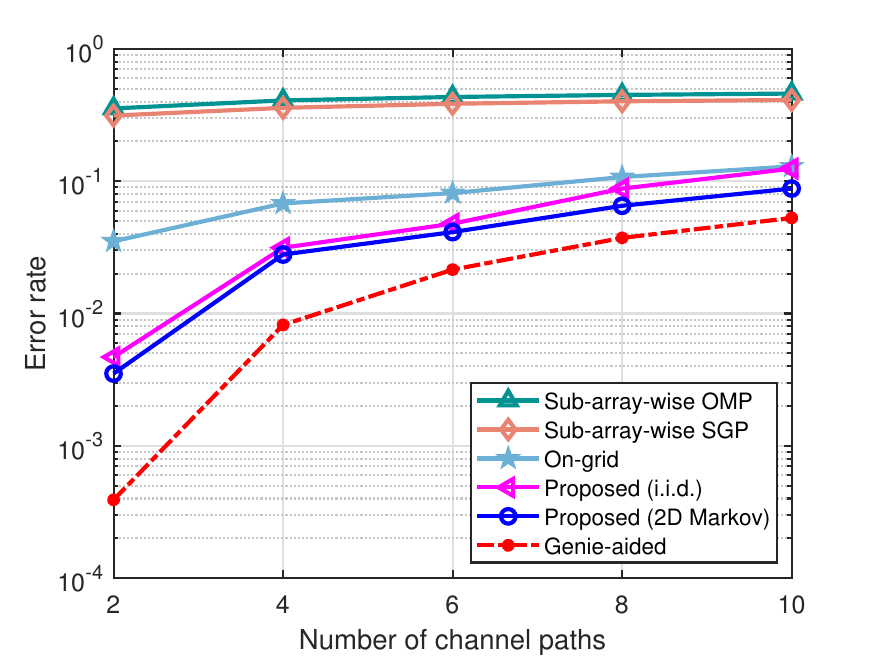}
\par\end{center}
\begin{center}
\vspace{-10mm}
\par\end{center}
\caption{\textcolor{blue}{\label{fig:Error_rate_nPath}}Error rate of VR detection
versus the number of channel paths when $\textrm{SNR}=-4\ \textrm{dB}$.}
\end{minipage}
\end{figure*}
\begin{figure*}[t]
\centering{}%
\begin{minipage}[t]{0.45\textwidth}%
\begin{center}
\includegraphics[clip,width=65mm]{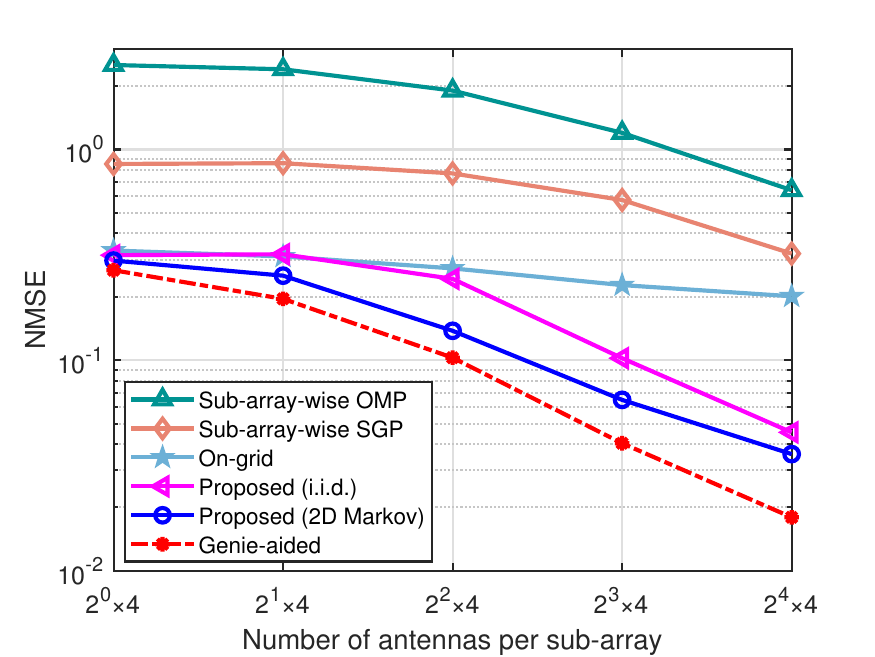}
\par\end{center}
\begin{center}
\vspace{-10mm}
\par\end{center}
\caption{\textcolor{blue}{\label{fig:NMSE_Nx}}NMSE of channel estimation versus
the number of antennas per sub-array. We set $\textrm{SNR}=-4\ \textrm{dB}$
and $N_{z}=4$.}
\end{minipage}\hfill{}%
\begin{minipage}[t]{0.45\textwidth}%
\begin{center}
\includegraphics[clip,width=65mm]{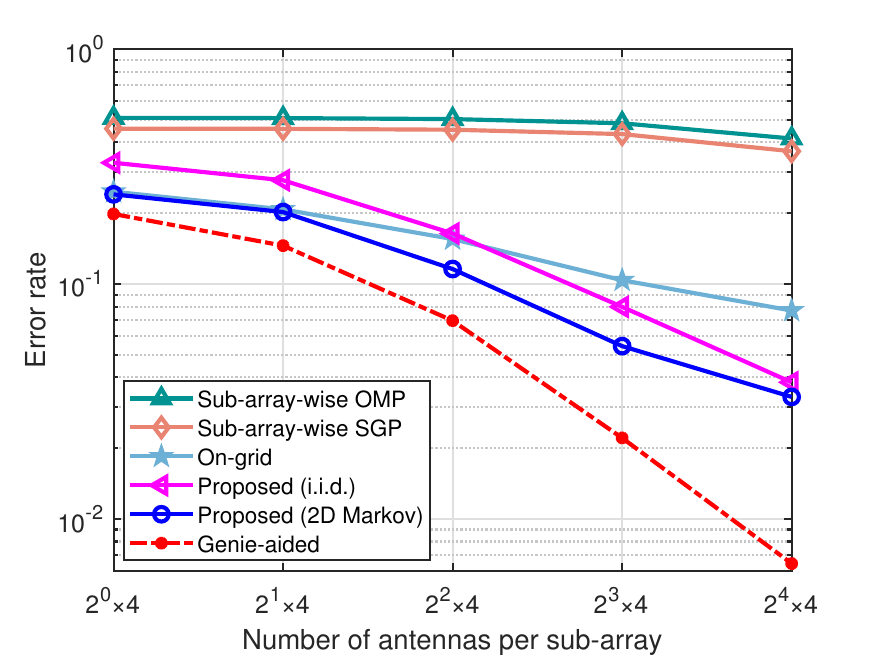}
\par\end{center}
\begin{center}
\vspace{-10mm}
\par\end{center}
\caption{\textcolor{blue}{\label{fig:Error_rate_Nx}}Error rate of VR detection
versus the number of antennas per sub-array. We set $\textrm{SNR}=-4\ \textrm{dB}$
and $N_{z}=4$.}
\end{minipage}
\end{figure*}

\subsection{Impact of SNR}

In Fig. \ref{fig:NMSE_SNR} and Fig. \ref{fig:Error_rate_SNR}, we
evaluate the performance of channel estimation and VR detection against
SNR, respectively. As can be seen, the performance of all methods
improves as the SNR increases. The sub-array-wise methods works poorly
since they ignore the fact that sub-channels share some common scatterers
as well as the associated channel parameters. Besides, the proposed
algorithm based on the dynamic grid has a significant performance
gain over the on-grid method, which verifies that the dynamic polar-domain
grid greatly improves the performance. Furthermore, in the low SNR
regions ($-10\ \textrm{dB}\sim-4\ \textrm{dB}$), the proposed algorithm
with the 2D Markov prior works better than the proposed algorithm
with the i.i.d. prior, which indicates that 2D Markov model can capture
the the 2D clustered sparse structure well. However, the performance
gain disappears in the high SNR regions. This is because the observation
information is enough to estimate the channel and VRs accurately when
the SNR is high. In this case, the prior information contributes little
to the performance of the algorithm. Finally, the performance of the
proposed algorithm with the 2D Markov prior is close to the genie-aided
method, which indicates that the dynamic grid parameters can be refined
well via gradient ascent. 

\subsection{Impact of Number of Channel Paths}

In Fig. \ref{fig:NMSE_nPath} and Fig. \ref{fig:Error_rate_nPath},
we focus on how the number of channel paths affects the performance
of different methods when $\textrm{SNR}=-4\ \textrm{dB}$. We vary
the number of channel paths from $L=2$ to $L=10$. There are more
non-zero channel parameters that need to be estimated when the number
of paths is larger. As a result, the performance of the methods decreases
gradually as the number of paths increases. Again, we find that the
proposed algorithm with the 2D Markov prior outperforms other baseline
methods. In addition, the performance gap between the proposed algorithm
with the 2D Markov prior and the i.i.d. prior is more obvious when
the number of paths is large. In this case, the 2D Markov prior information
plays a key role in the algorithm.

\subsection{Impact of Number of Antennas Per Sub-array}

Fig. \ref{fig:NMSE_Nx} and Fig. \ref{fig:Error_rate_Nx} plot the
NMSE of channel estimation and error rate of VR detection against
the number of antennas per sub-array, respectively. We keep $N_{z}=4$
and vary $N_{x}$ from $2^{0}$ to $2^{4}$. Note that the sub-array
is spatial stationary when $N_{x}\leq16$ and $N_{z}\leq4$, and a
$16\times4$ sub-UPA is the largest unit keeping spatial stationary.
It can be seen that with the increase in $N_{x}$, the NMSE and VR
error rate of the methods reduced. This is because different antennas
in a same sub-array share the same VRs. As a result, as $N_{x}$ increases,
the sub-array-specific sparsity increases, which the proposed method
fully exploit. We also notice that for all sub-array sizes, the proposed
algorithm with the 2D Markov prior works better than baselines. 

\section{Conclusions}

We propose a joint VR detection and channel estimation method for
XL-MIMO systems. Based on the polar-domain sparse representation of
the XL-MIMO channel, we use a hierarchical sparse prior model to capture
the sparsity of the channel vector. Besides, a 2D Markov model is
designed to fully exploit the 2D clustered sparsity of VRs. Based
on these, the considered problem is formulated as a MAP estimation
problem. A novel alternating MAP framework is developed to solve the
problem by combining the IF-VBI estimator, structured EP algorithm,
and gradient ascent approach. Three basic modules of the proposed
alternating MAP framework work alternatively to estimate the polar-domain
channel vector, detect the VRs, and refine the dynamic grid parameters.
Simulations verify that our proposed method outperforms baselines
in terms of both channel estimation and VR detection. 

In future work, we will extend the proposed method to more challenging
scenarios, such as channel tracking with multiple time slots, multiple
users with pilot contamination, and the receiver with a hybrid beamforming
(HBF) architecture.

\bibliographystyle{IEEEtran}
\bibliography{NF}

\end{document}